\documentclass[preprint,12pt]{revtex4-1}

\usepackage[brazil, english]{babel}
\usepackage[utf8]{inputenc}
\usepackage{graphicx}
\usepackage{epsfig}
\usepackage{amssymb}
\usepackage{amsmath} 
\usepackage{soul}
\usepackage[unicode=true,bookmarks=false,breaklinks=false,pdfborder={0 0 1},colorlinks=true]
 {hyperref}
\hypersetup{
 citecolor=blue,linkcolor=blue,urlcolor=blue}
\usepackage[mathscr]{euscript}
\usepackage{mathrsfs}
\usepackage{array}
\usepackage{slashed}
\usepackage{dsfont}
\usepackage[normalem]{ulem}
\usepackage{verbatim}%pacote para usar o \begin comment adicionado por Nathan 

\begin{document}

%\date{}
\title{Fluctuation and Dissipation from a Deformed String/Gauge Duality Model}
\author{Nathan G. Caldeira$^{1,}$}
\email[Eletronic address: ]{nathangomesc@hotmail.com}
\author{Eduardo Folco Capossoli$^{1,2,}$}
\email[Eletronic address: ]{eduardo\_capossoli@cp2.g12.br}
\author{Carlos A. D. Zarro$^{1,}$}
\email[Eletronic address: ]{carlos.zarro@if.ufrj.br}
\author{Henrique Boschi-Filho$^{1,}$}
\email[Eletronic address: ]{boschi@if.ufrj.br}  
\affiliation{$^1$Instituto de F\'{\i}sica, Universidade Federal do Rio de Janeiro, 21.941-972 - Rio de Janeiro-RJ - Brazil \\ 
 $^2$Departamento de F\'{\i}sica and Mestrado Profissional em Práticas de Educa\c{c}\~{a}o B\'{a}sica (MPPEB), 
 Col\'egio Pedro II, 20.921-903 - Rio de Janeiro-RJ - Brazil\\
 %$^3$Instituto de F\'{\i}sica, Universidade Federal do Rio de Janeiro, 21.941-972 - Rio de Janeiro-RJ - Brazil
}

\begin{abstract}
Using a Lorentz invariant deformed string/gauge duality model at finite temperature we calculate the thermal fluctuation and the corresponding linear response, verifying the fluctuation-dissipation theorem. The deformed AdS$_5$ is constructed by the insertion of an exponential factor $\exp(k/r^2)$ in the metric. We also compute the string energy and the mean square displacement in order to investigate the ballistic and diffusive regimes. Furthermore we have studied the dissipation and the linear response in the zero temperature scenario.

%In this work we calculate the static limit of the energy for a quark-antiquark pair from the Nambu-Goto action using a holographic approach with a deformed AdS space. From this energy we 
%derive the Cornell potential for the quark-antiquark interaction. We also find a range of values for our parameters which fits exactly the Cornell potential parameters.  
\end{abstract}

%\pacs{11.25.Wx, 11.25.Tq, 12.38.Aw, 12.39.Mk}

\maketitle

\section{Introduction}\label{sec:Introduction}
\noindent
\begin{comment}
It is a universal phenomenon that a particle in a thermal bath, if isolated, will experience two different 

have its motion described by two mechanisms. First it will experience some dissipation when      
\end{comment}

Brownian motion is a rather universal  phenomenon first observed  occurring for pollen grains suspended in liquids \cite{R:Brown}. Such particles in this environment exhibit an apparently random motion. The description of this kind of system is given by the Langevin equation what shapes the force acting on the particle as being composed by a dissipative part and a component related with the random fluctuations \cite{Langevin:P}. 

Usually the Langevin equation is written as \cite{Kubo1966} 
\begin{equation}
\label{Langevin}
    \frac{d \vec{p}}{dt}=-\gamma \vec{p} +\vec{R}(t)+\vec{F}(t)\,,
\end{equation}
\noindent 
where the first term on right side quantifies the dissipative force acting on the particle and $\gamma$ is the (constant) friction coefficient. The second part, $\vec{R}(t)$, is related with the random fluctuations affecting the motion of the particle. It is a stochastic variable with zero average and white noise: 
\begin{align}
\label{Correlation}
     \langle R_{i}(t)\rangle=0 && \langle R_{i}(t)R_{j}(t')\rangle=\kappa\delta_{ij}\delta(t-t')\,. 
\end{align}
The third part of the Langevin equation \eqref{Langevin} is a possible external force involved. It can be generalized considering the friction force dependent on the history of the motion and a more general correlation for the noise. For a more complete discussion see, for example, \cite{toda1991statistical}.     

A cornerstone on the study of the Brownian motion and statistical systems in general was given by Kubo \cite{Kubo1966} in which he analysed the Langevin equation and  the linear response theory. One of his most interesting results is the well-known fluctuation-dissipation theorem (FDT) which  relates quantities linked with fluctuation in equilibrium state with others concerning the dissipation process. It is a major result because it puts together those two important parts of the description of a thermal system. In fact any system in a thermal bath will experience those two effects. What this theorem tells us is that such processes are not independent (see also \cite{toda1991statistical}).

This theorem applies to a large class of systems as,  for instance, in the description of the Johnson-Nyquist noise in electric circuits where thermal fluctuations of the electrons give rise to potential differences between the components \cite{PhysRev.32.97, PhysRev.32.110}.  Arguments related with the FDT are also used in the analysis of the dynamical Casimir effect. In that case one can calculate the Casimir dissipative force on objects in motion from the fluctuations of the force acting on them at rest. One example is the computation of a dissipative force on a  perfectly reflecting moving sphere in Ref.  \cite{MaiaNeto:1993zz}. Another interesting application of FDT is, for instance, in the theory of lasers  involving the admittance of optical cavities and their absorption of thermal radiation  \cite{PhysRevA.18.659,PhysRevLett.93.213905,doi:10.1002/er.1607}.

The AdS/CFT correspondence  was formulated as a duality between a $IIB$ superstring theory living in AdS$_5\times S^5$ space and a superconformal ${\cal N} = 4$ Yang-Mills theory, with symmetry group $SU(N \to \infty)$, defined in a Minkowski spacetime on the boundary of the AdS$_5$ space \cite{Maldacena:1997re,Gubser:1998bc,Witten:1998qj,Witten:1998zw,Aharony:1999ti}. One of the main achievements of the AdS/CFT correspondence is to describe a weak coupled theory living in AdS$_5$ space bulk as a strongly coupled theory on the boundary. Such a duality is appropriate to deal with  thermodynamic features of some systems as the  quark-gluon plasma (QGP) \cite{Policastro:2001yc, CasalderreySolana:2011us}. QGP is a very useful system for our purpose since the hadronic matter at extremely high temperatures and densities seems to exhibit random walks due to their collision with each other just like a Brownian motion. In other words, one can use QGP at finite temperature within holographic approaches to study Brownian motion, quantum fluctuations, dissipation, linear response, etc. 

Many works were done in this direction within holographic contexts, for example, dealing with Brownian motion \cite{deBoer:2008gu, Son:2009vu, Atmaja:2010uu, Chakrabortty:2013kra, Sadeghi:2013jja, Banerjee:2013rca, Banerjee:2015vmo, Chakrabarty:2019aeu}, fluctuation or dissipation  \cite{Tong:2012nf,  Edalati:2012tc, Kiritsis:2012ta, Fischler:2014tka, Roychowdhury:2015mta, Banerjee:2015fed, Giataganas:2018ekx}, drag forces \cite{Gubser:2006bz, Gubser:2006qh, Kiritsis:2013iba, Andreev:2017bvr, Andreev:2018emc, Bena:2019wcn, Diles:2019jkw, Tahery:2020tub} and related topics \cite{Kinar:1999xu, Gursoy:2010aa, Giataganas:2013zaa, Giataganas:2013hwa, Sadeghi:2014lha, Dudal:2014jfa, Dudal:2018rki}. In particular, de Boer {\sl et al.} studied the Brownian motion in a CFT described from AdS black holes  \cite{deBoer:2008gu}. Tong and Wong \cite{Tong:2012nf} discussed quantum fluctuations in a Lifschitz spacetime breaking Lorentz symmetry. Edalati, Pedraza and Tangarife Garcia \cite{Edalati:2012tc} considered a hyperscale violation in quantum and thermal fluctuations. These works were extended by Giataganas, Lee and Yeh  \cite{Giataganas:2018ekx} dealing  with  Brownian motion, fluctuation, dissipation in a general context for a polynomial metric. 

In order to describe fluctuations in QCD-like theories from the AdS/CFT correspondence one has to introduce an infrared scale breaking conformal invariance. In hadronic physics there are basically two approaches to do that known as top-down \cite{Klebanov:1998hh, Klebanov:1999tb, Klebanov:2000hb, Maldacena:2000mw, Maldacena:2000yy, Sakai:2004cn, Sakai:2005yt} and bottom-up  \cite{Polchinski:2001tt,Polchinski:2002jw,BoschiFilho:2002vd,BoschiFilho:2002ta,BoschiFilho:2005yh,Capossoli:2013kb,Rodrigues:2016cdb, Karch:2006pv, Colangelo:2007pt,Li:2013oda,Capossoli:2015ywa,Capossoli:2016kcr,Capossoli:2016ydo, FolcoCapossoli:2016ejd, MarinhoRodrigues:2020ssq}. In the bottom-up approach, the first proposal is known as the hardwall model which introduces a hard cut off in AdS space \cite{Polchinski:2001tt,Polchinski:2002jw,BoschiFilho:2002vd,BoschiFilho:2002ta,BoschiFilho:2005yh,Capossoli:2013kb,Rodrigues:2016cdb}.  The second proposal is known as the softwall model and it introduces a dilaton field in the action playing the role of soft cut off \cite{Karch:2006pv, Colangelo:2007pt,Li:2013oda,Capossoli:2015ywa,Capossoli:2016kcr,Capossoli:2016ydo, FolcoCapossoli:2016ejd, MarinhoRodrigues:2020ssq}. An alternative for the softwall model  is to introduce a warp factor deformation in the metric instead of the dilaton in the action. Within this approach one can calculate  quark-antiquark potential at zero and finite temperature, hadronic spectra, etc   \cite{Andreev:2006vy, Andreev:2006ct, Wang:2009wx, Afonin:2012jn, Rinaldi:2017wdn, Bruni:2018dqm, Afonin:2018era, Diles:2018wbe, FolcoCapossoli:2019imm,Rinaldi:2020ssz}. 

Then, one can use some of these ideas from the AdS/CFT approach to hadronic physics in order to investigate Brownian motion, fluctuations, dissipation, etc.  For instance, Ref.  \cite{Dudal:2018rki} studied heavy quark diffusion in the presence of a magnetic field  introducing an exponential factor in the Nambu-Goto action. In Ref. \cite{Tahery:2020tub} they calculated the drag force in a moving heavy quark using the deformed AdS space proposed in \cite{Andreev:2006vy}. 

The main goal of this work is to study zero and finite temperature string fluctuations using a deformed AdS space with the introduction of an exponential factor $\exp{k/r^2}$ in the metric, motivated by the success of this approach to hadronic physics \cite{Andreev:2006vy, Andreev:2006ct, Wang:2009wx, Afonin:2012jn, Rinaldi:2017wdn, Bruni:2018dqm, Afonin:2018era, Diles:2018wbe, FolcoCapossoli:2019imm,Rinaldi:2020ssz}. We calculate thermal fluctuations, the admittance from linear response,  two-point functions and show explicitly that the fluctuation-dissipation theorem holds in this setup.  Notice that the analysis of Ref. \cite{Giataganas:2018ekx} can be applicable up to certain orders also for the boundary and horizon expansions of generic form metric fields. We complete our study with the zero temperature response function calculating the corresponding admittance.

This work is organized as follows. In Section \ref{gravset} we introduce our geometric setup at finite temperature, calculate the energy of the string, find the equations of motion and their solutions in different regions in the deformed AdS black hole space and impose matching conditions between these solutions. In Section \ref{FDT} we compute the admittance through the linear response theory, the thermal two-point functions, the mean square displacement. From this result we obtain the ballistic and diffusive regimes of the Brownian motion of the particle described holographically by the end of the open string. From the relation between the imaginary part of the admittance and the two-point functions we verify the fluctuation-dissipation theorem.  In Section \ref{Tzero}, we reconsider the previous setup for the case of zero temperature and calculate the corresponding admittance from the linear response theory. Finally, in Section \ref{sec:conclusions}, we present our last comments and conclusions.

%%%%%%%%%%%%%%%%%%%%%%%%%%%%%%%%%%%%%%%%%%%%%%%%%
%%%%%%%%%%%%%%%%%%%%%%%%%%%%%%%%%%%%%%%%%%%%%%%%%

\section{String/Gauge setup at finite temperature}\label{gravset}

In this section, we are going to introduce our string/gauge setup at finite temperature to investigate the holographic Brownian motion. Since we are  interested in a Lorentz-invariant scenario, instead of a scaling  violation \cite{Tong:2012nf, Edalati:2012tc, Giataganas:2018ekx}, here 
the conformal invariance is broken by introducing an exponential factor in $AdS_5$ metric following ref. \cite{Andreev:2006ct}:
\begin{equation}\label{metrictemp}
    ds^2 = e^{\frac{k}{r^2}} \left[-r^{2}f(r)dt^{2}+{r^2}\left(\eta_{i j} dx^{i}dx^{j} \right) +\frac{dr^2}{r^{2}f(r)}\right],
\end{equation}
where $\eta_{i j} = {\rm diag} (-1, +1, +1 , +1)$, the AdS radius was set to 1, $r$ is holographic coordinate and $f(r)$ is called the horizon function which is given by:
\begin{equation}\label{HF}
    f(r)=\left(1-\frac{r_{h}^{4}}{r^{4}}\right)\,, 
\end{equation}
\noindent and $r_h$ is the horizon radius. In Refs. \cite{Andreev:2006vy, Andreev:2006ct, Wang:2009wx, Afonin:2012jn, Rinaldi:2017wdn, Bruni:2018dqm, Afonin:2018era, FolcoCapossoli:2019imm} this metric was used to study many aspects of holographic high energy physics. In these references, $k$ is a constant which can be related to  $\Lambda_{QCD}$. It is important to mention that the algebraic sign of $k$ is not a consensus in the literature (see for instance \cite{Karch:2006pv, Andreev:2006ct, Karch:2010eg}) and we will comment on this in further sections. The corresponding Hawking temperature is given by:
\begin{equation}
    T = \frac{K_H}{2 \pi} \sqrt{\frac{g_{tt}(r_h)}{g_{rr}(r_h)}},
\end{equation}{}
where $K_H$ is the surface gravity given by $K_H = (1/2) f'(r_h)$. So, for the metric \eqref{metrictemp} the temperature  is related to horizon radius:
\begin{equation} \label{thor}
    T = \frac{r_h}{\pi}.
\end{equation}
 One of the main features of our model is to get, at same time, the breaking of the conformal invariance and to be Lorentz invariant, such that we can obtain correctly the fluctuation-dissipation theorem. Besides such a deformation reproduces the $AdS_5$ space close to UV region $(r \to \infty)$.

\begin{figure}
	\centering
	\includegraphics[scale = 0.7]{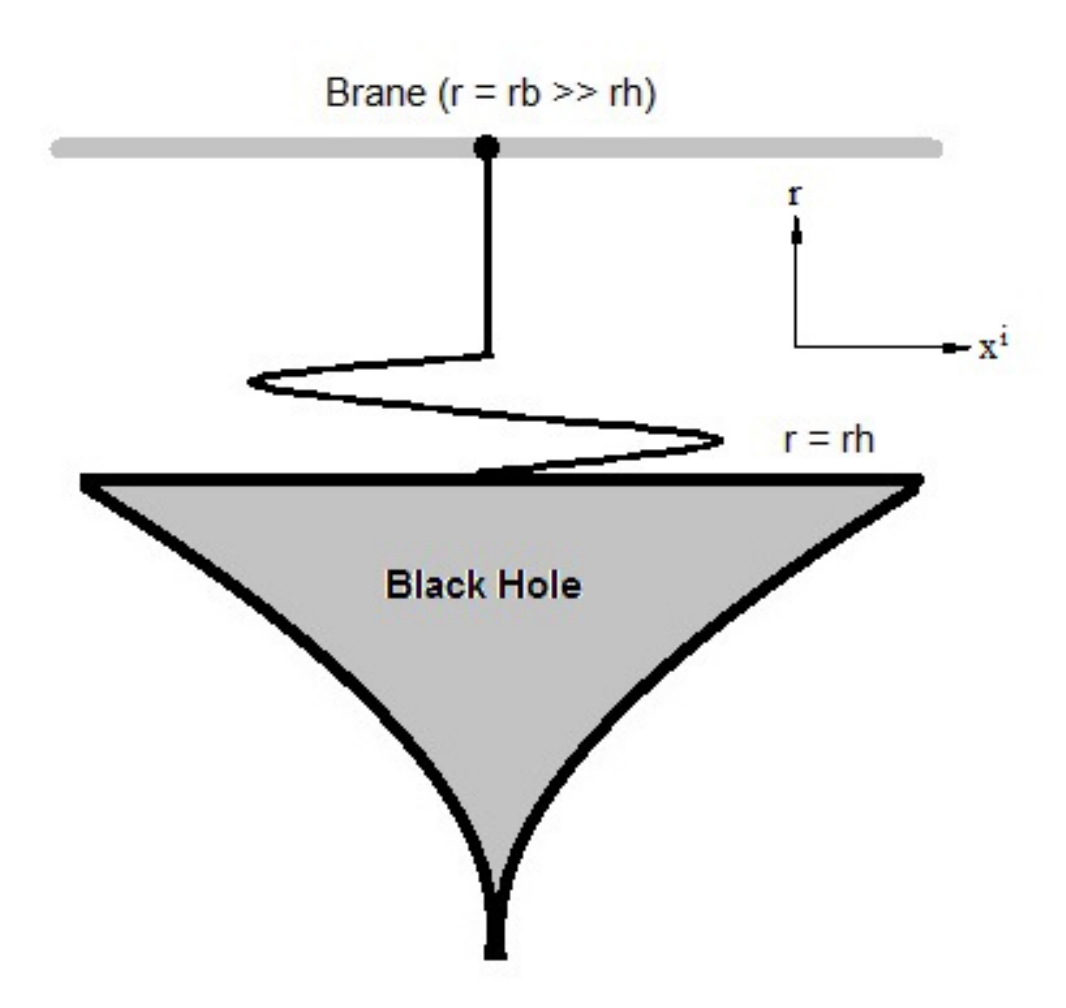}
	\caption{The string/gauge setup for the holographic Brownian motion.}
	\label{fig:f(z)}
\end{figure}

According to string/gauge duality a massive particle can be understood as the endpoint of an open string. This endpoint is attached to a probe brane located at $r=r_b$ close to the boundary $(r \to \infty)$. The string extends itself to entire bulk, hence, its other endpoint is placed at the IR region with $r \to r_h$, where $r_h$ is the horizon of the black hole, as can be seen in Fig. \ref{fig:f(z)}.

The Brownian motion of the massive particle at the brane is explained as the vibration of the string endpoint near the horizon which  interacts with the Hawking radiation. Once we established our geometric setup, the string dynamics is described by the Nambu-Goto action, so that:
\begin{equation}\label{ng}
S_{NG} = - \frac{1}{2 \pi \alpha'} \int d\tau d\sigma \sqrt{-\gamma}\, ,
\end{equation}
\noindent where $\alpha'$ is the string tension, $\gamma = {\rm det} (\gamma_{\alpha \beta})$ and $\gamma_{\alpha \beta} = g_{mn} \partial_{\alpha}X^m \partial_{\beta}X^n $ is the induced metric on the worldsheet with $m,n = 0, 1, 2, 3, 5$.

As done in Refs. \cite{Edalati:2012tc, Giataganas:2018ekx} we also choose a static gauge, where $t = \tau$, $r = \sigma$ and $X= X(\tau, \sigma)$. By using the metric, Eq. \eqref{metrictemp}, and expanding the Nambu-Goto action, Eq. \eqref{ng}, in order to keep the quadratic terms $\dot{X}^2$, $X'^2$, we get: 
\begin{equation}\label{ngapprox}
S_{NG} \approx - \frac{1}{4 \pi \alpha'} \int d\tau d\sigma \left[ \;\dot{X}^2 \frac{e^{\frac{ k}{r^2}}}{f(r)}-X'^2 r^4 f(r) e^{\frac{k}{r^2}} \right]\,,
\end{equation}
\noindent where $\dot{X}=\partial_{\tau =t} X$ and $X'=\partial_{\sigma =r} X$.

Following \cite{Giataganas:2018ekx,Tong:2012nf}, we can compute the energy to create the string described above as  
\begin{equation}
    E=\frac{1}{2\pi \alpha'}\int_{r_{h}}^{r_{b}} \sqrt{-g_{00}g_{rr}}= \frac{1}{2\pi \alpha'}\int_{r_{h}}^{r_{b}} e^{\frac{k}{r^{2}}}.
\end{equation}
For $k>0$, one finds that
\begin{equation}\label{eq:Estringk>0}
 E_{k>0}=\frac{1}{2 \pi  \alpha' }\left\{r_{b} e^{\frac{k}{r_{b}^2}}-r_{h} e^{\frac{k}{r_{h}^{2}}}+\sqrt{\pi k} \left[\text{Erfi}\left(\frac{\sqrt{k}}{r_h}\right)-\text{Erfi}\left(\frac{\sqrt{k}}{r_b}\right)\right]\right\},
 \end{equation}
where $\text{Erfi}$ is the imaginary error function, defined as $\text{Erfi}(z)=\text{Erf}(iz)/i$ where $\text{Erf}(z)$ is the error function given by $\text{Erf}(z)=(2/\sqrt{\pi})\int_{0}^{z}e^{-t^{2}}dt$ \cite{abramowitz+stegun}. The energy for $k<0$ reads:
\begin{equation}\label{eq:Estringk<0}
 E_{k<0}=\frac{1}{2 \pi  \alpha' }\left\{r_{b} e^{\frac{-|k|}{r_{b}^2}}-r_{h} e^{\frac{-|k|}{r_{h}^{2}}}+\sqrt{\pi |k|} \left[\text{Erf}\left(\frac{\sqrt{|k|}}{r_h}\right)-\text{Erf}\left(\frac{\sqrt{|k|}}{r_b}\right)\right]\right\}.
 \end{equation}
The AdS limit can be obtained for $|k|\ll r_{h}\ll r_{b}$, for both signs of $k$ as given by Eqs. \eqref{eq:Estringk>0} and \eqref{eq:Estringk<0} so that: 
\begin{equation}\label{eq:EAdS}
 E_{AdS}=\frac{(r_{b} -r_{h})}{2 \pi  \alpha'}\approx \frac{r_{b}}{2\pi\alpha'},
 \end{equation}
and the energy of the string is proportional to its length which is approximated by $r_b$ as expected. 

%\begin{figure}[!ht]
%	\centering
%	\includegraphics[scale = 0.6]{Ekpos.pdf}
%	\caption{String energy, E, Vs. $r_b$ for a fixed $r_h$ and positive $k = 1$ GeV$^2$. The red dot represents the maximum energy for $r_b \to r_h$. }
%	\label{epos}
%\end{figure}
%
%\begin{figure}[!ht]
%	\centering
%	\includegraphics[scale = 0.6]{Ekneg.pdf}
%	\caption{String energy, E, Vs. $r_b$ for a fixed $r_h$ and negative $k = -1$ GeV$^2$. The red dot represents the minimum energy for $r_b \to r_h$.}
%	\label{eneg}
%\end{figure}

The equation of motion for the string described by $X(t,r)$ can be derived from the approximate Nambu-Goto action, Eq. \eqref{ngapprox}: 
\begin{equation}\label{eq:EOM-NG}
    \frac{\partial }{\partial r}\left(r^{4}f(r)e^{\frac{k}{r^{2}}}X'(r,t)\right)-\frac{e^{\frac{k}{r^{2}}}}{f(r)}\ddot{X}(t,r)=0.
\end{equation}
Performing the following ansatz $X(t,r)=e^{i\omega t}h_{\omega}(r)$, one gets:
\begin{equation}\label{ansatz}
    \frac{d }{d r}\left(r^{4}f(r)e^{\frac{k}{r^{2}}}h_{\omega}'\right)+\frac{\omega^{2}e^{\frac{k}{r^{2}}}}{f(r)}h_{\omega}=0.
\end{equation}
Changing the variable $r$ to the tortoise coordinate $r_{*}$ defined as
\begin{equation}\label{tartaruga}
    r_{*}=\int\frac{dr}{r^{2}f(r)}=\frac{1}{4r_ {h}} \log \left(\frac{r-r_ {h}}{r+r_ {h}}\right)  + \frac{1}{2r_{h}} \tan^{-1}\left (\frac {r}{r_ {h}} \right),
\end{equation}
one obtains
\begin{equation}
    r^2 e^{\frac{k}{r^{2}}}\frac{d^{2}h_{\omega}}{dr_{*}^{2}}+\frac{d}{dr_{*}}\left(r^2 e^{\frac{k}{r^{2}}}\right)\frac{d h_ {\omega}}{dr_{*}}+\omega ^2  r^2 e^{\frac{k}{r^2}}h_{\omega}=0,
\end{equation}
where $r=r(r_{*})$ and $h_{\omega}=h_{\omega}(r_{*})$. The following substitution
\begin{equation}\label{hpsi}
    h_{\omega}=e^{B(r_{*})}\psi(r_*),\,\,\, 
\end{equation}
 where $B(r_{*})= -{k}/({2 r^2})-\log (r)$ gives the following Schr\"{o}dinger-like equation:
\begin{equation}\label{sch}
    \frac{d^{2}\psi(r_{*})}{dr_{*}^{2}}+\left(\omega^{2}-V(r)\right)\psi(r_{*})=0,
\end{equation}
where
\begin{equation}\label{potsch}
    V(r)=2 r^2-k +\frac{k^2}{r^2}-\frac{2k r_{h}^4}{r^4}-\left(\frac{2 k^{2} r_{h}^{4}+ 2r_ {h}^{8}}{r^6}\right) +\frac{3 k r_{h}^8}{r^8}+\frac{k^2  r_{h}^8}{r^{10}}.
\end{equation}
Notice that $V(r_{h})=0$, as expected. Nearby the horizon, the potential can be expanded in Taylor series as:
\begin{equation}
    V(r)\approx 16\left(-\frac{k}{r_{h}}+r_{h}\right)(r-r_{h}).
\end{equation}
The Schr\"{o}dinger-like equation \eqref{potsch} cannot be analytically solved, hence one seeks for solutions within certain regions. For our purposes we will choose three regions: \textbf{A}, \textbf{B}, \textbf{C} and explore their solutions.

The first region, dubbed as {\bf A}, is nearby the event horizon, $i.e.$ $r\approx r_{h}$. In this case the, $V(r)\ll\omega^{2}$, and the Schr\"{o}dinger-like equation reads
\begin{equation}
    \frac{d^{2}\psi(r_{*})}{dr_{*}^{2}}+\omega^{2}\psi(r_{*})=0,
\end{equation}
which has the ingoing solution
\begin{equation}\label{solA}
    \psi(r_{*}) = A_1 e^{-i\omega r_*}.
\end{equation}
Nearby the horizon ($r\approx r_{h}$), we can assume that for low frequencies we have $\omega r_* \ll 1$. Then one can expand Eq. \eqref{solA} as: 
\begin{equation}\label{psir*AA}
    \psi(r_{*}) = A_1  -i A_1 \omega r_* .
\end{equation}
Using this equation and Eq. \eqref{hpsi}, we can compute $h_{\omega}(r_{*})$ in this region:
\begin{equation}\label{hAr*}
h_{\omega}^{A}(r_*)=\frac{e^{-\frac{k}{2 r^2_h}}}{r_h}\left(A_{1}- i\omega A_1 r_* \right),
\end{equation}
where $r_*$ is given by \eqref{tartaruga}. In the limit $r\approx r_{h}$, we find
\begin{equation}\label{rstarA}
   r_* = \frac{1}{4 r_h} \log(r - r_h) - \frac{\log (2 r_h)}{4 r_h} + \frac{\pi}{8 r_h}. 
\end{equation}{}
Substituting this equation into  Eq. \eqref{hAr*} we get
\begin{equation}\label{HA}
 h_{\omega}^{A}(r)= \frac{e^{-\frac{k}{2 r^2_h}}}{r_h}\left(\tilde{A}_{1} - \frac{i\omega A_1}{4r_ {h}}\log(r - r_h) \right),
\end{equation}
\noindent where 
\begin{equation}\label{atil}
    \tilde{A}_{1} =A_{1}+\frac{i\omega A_1}{4r_ {h}} \log(2r_{h})-\frac{i \pi \omega A_1}{8r_ {h}}.
\end{equation}

Following Ref. \cite{deBoer:2008gu}, one has to impose a regularization procedure by introducing a cutoff at $r=r_{h}+\epsilon$ nearby the horizon, $i.e.$ $\epsilon \ll 1$. The complete solution in this region comprises the ingoing and outgoing modes:
\begin{equation}
    f^{A}_{\omega}(t,r)=A_{\omega}\left[\frac{e^{\frac{k}{r^{2}}}}{r}e^{-i\omega r_*}+B_{\omega}\frac{e^{\frac{k}{r^{2}}}}{r}e^{i\omega r_*}\right]e^{-i\omega t}.
\end{equation}
Imposing the Neumann boundary condition at $r=r_{h}+\epsilon$, one finds
\begin{equation}
    \left.\frac{d f^{A}_{\omega}}{dr}\right|_{r=r_{h}+\varepsilon}=0 \Leftrightarrow B_{\omega}= e^{2 i \omega  \left(\frac{\pi }{8 h}-\frac{\log (2 r_h)}{4 r_h}\right) } e^{\frac{-i\omega }{2 r_h} \log\left(\frac{1}{\epsilon}\right)}.
\end{equation}
The above condition implies that the possible frequencies are now discrete:
\begin{equation}\label{eq:discretenesscondition}
    \Delta \omega = \frac{4\pi r_h}{\log\left(\frac{1}{\varepsilon}\right)}.
\end{equation}

The region {\bf B} corresponds to  $\omega^2 \ll V(r)$, which implies $\omega^2 \ll f(r)$. In this regime, Eq. \eqref{ansatz} has the following form:
\begin{equation}\label{reg2}
    \frac{d h_{\omega}}{d r} = \frac{B_{1}}{r^4 f(r) e^{k/r^2}}, 
\end{equation}
\noindent where $f(r)$ is given by Eq. \eqref{HF} and $B_{1}$ is a constant. This equation can be integrated to
\begin{equation}\label{hBb}
    h_{\omega}^{B}(r)=B_1 \int^{r}\frac{e^{-k/r^{2}}}{r^{4}-r^{4}_{h}}\;dr + B_2,
\end{equation}
where $B_{1}$ and $B_{2}$ are integration constants. In the IR limit, $i.e.$ for $r\sim r_{h}$, one has
\begin{equation}
    r^{4}-r^{4}_{h}=(r-r_{h})(r^{3}+r_{h} r^{2}+r_{h}^{2}r+r_{h}^{3}),
\end{equation}
hence for $r\sim r_{h}$, our integral can be approximated by
\begin{equation}
    h_{\omega(IR)}^{B}(r)\approx B_1 \frac{e^{-k/r^{2}_{h}}}{4r_{h}^{3}} \int^{r}\frac{dr}{r-r_{h}}\;dr + B_2\approx B_1 \frac{e^{-k/r^{2}_{h}}}{4 r_{h}^{3}}  \log(r-r_{h}) + B_{1}\frac{e^{-k/r^{2}_{h}}}{4r_{h}^{3}} b + B_2, \label{hBIR}
\end{equation}
where $b$ is an integration constant. Now, we are going to obtain the UV limit in region {\bf B}. In this case, the integral of Eq. \eqref{hBb}, in the limit $r\gg r_{h}$, becomes
\begin{equation}\label{hB}
    h_{\omega (UV)}^{B}(r)\approx B_1 \int^{r}_{\infty}\frac{e^{-k/r^{2}}}{r^{4}}\;dr + B_2\approx B_1 \int^{r}_{\infty} \frac{dr}{r^4} + B_2 = - \frac{B_1}{3 r^3} + B_2.
\end{equation}
The third region, {\bf C}, that we will analyze corresponds to $r \to \infty$ meaning that the horizon function $f(r) \to 1$. In this case, Eq. \eqref{ansatz} has the following solution:
\begin{equation}
h^C_{\omega}(r)=C_1\Phi\left(\frac{\omega^{2}}{4k},-\frac{1}{2},-\frac{k}{r^{2}}\right)+ C_2 \frac{(-k)^{3/2}}{r^3}\Phi\left(\frac{3}{2} + \frac{\omega^{2}}{4k},-\frac{5}{2},-\frac{k}{r^{2}}\right), 
\end{equation}
\noindent where $\Phi(a,b,c)$ is the confluent hypergeometric function of the first kind \cite{abramowitz+stegun}. In the limit $r \to \infty$ its  asymptotic expression is given by:
\begin{equation}\label{HwC}
h^C_{\omega}(r)= C_1+ \frac{C_1 \omega ^2}{2 r^2} + \frac{C_2 (-k)^{3/2}}{r^3} +O\left(\left(\frac{1}{r}\right)^4\right). 
\end{equation}{}
For small frequencies $\omega \to 0$, it reads
\begin{equation}
    h^C_{\omega}(r)= C_1+  \frac{C_2 (-k)^{3/2}}{r^3}.
\end{equation}

In order to relate these  constants, one has to connect the solutions found for each region \textbf{A}, \textbf{B} and \textbf{C}. Let us start matching the solutions in region \textbf{A} and the IR limit of region \textbf{B},  meaning $  h_{\omega}^{A}(r)=h_{\omega(IR)}^{B}(r)$, so that:
\begin{eqnarray}
    \frac{e^{-\frac{k}{2 r^2_h}}}{r_h}\left(\tilde{A}_{1} - \frac{i\omega A_1}{4r_ {h}}\log(r - r_h) \right) =  B_1 \frac{e^{-k/r^{2}_{h}}}{4r_{h}^{3}} \log(r-r_{h}) + B_{1}\frac{e^{-k/r^{2}_{h}}}{4r_{h}^{3}} b + B_2\,,
\end{eqnarray}{}
then one gets:
\begin{equation}
    \tilde{A}_{1} \frac{e^{-\frac{k}{2 r^2_h}}}{r_h} = B_{1}\frac{e^{-k/r^{2}_{h}}}{4r_{h}^{3}} b + B_2
\end{equation}{}
and
\begin{equation}
     B_1 = -i A_1 r_h \omega  e^{\frac{k}{2 r_h^2}}
\end{equation}

Now, the matching between the UV limit for region \textbf{B} and region \textbf{C} implies that $ h_{\omega(UV)}^{B}(r) =  h_{\omega}^{C}(r)$, therefore:
\begin{equation}
 - \frac{B_1}{3 r^3} + B_2 = C_1+  \frac{C_2 (-k)^{3/2}}{r^3},
\end{equation}{}
then one gets:
\begin{equation}
  C_1 = B_2 =  \tilde{A}_{1} \frac{e^{-\frac{k}{2 r^2_h}}}{r_h}  - B_{1}\frac{e^{-\frac{k}{r^{2}_{h}}}}{4r_{h}^{3}} b = \tilde{A}_{1} \frac{e^{-\frac{k}{2 r^2_h}}}{r_h}  + \frac{i A_1  \omega}{4r_{h}^{2}}  e^{-\frac{k}{2 r_h^2}} b 
\end{equation}
and
\begin{equation}
     C_2 = - \frac{1}{3} \left(-i A_1 r_{h} \omega  e^{\frac{k}{2 r^2_{h}}}\right) \frac{1}{(-k)^{3/2}}\,.
\end{equation}{}

Substituting these constants in Eq. \eqref{HwC} one gets:
\begin{equation}\label{hcfinal}
h^{C}_{\omega}(r)= \tilde{A}_{1} \frac{e^{-\frac{k}{2 r^2_h}}}{r_h} + \frac{i A_1  \omega}{4r_{h}^{2}}  e^{-\frac{k}{2 r_h^2}} b + \frac{\tilde{A}_{1} \frac{e^{-\frac{k}{2 r^2_h}}}{r_h} \omega ^2}{2 r^2} + \frac{ - \frac{1}{3} \left(-i A_1 r_h \omega  e^{\frac{k}{2 r_h^2}}\right) }{r^{3}},
\end{equation}
where 
\begin{equation}\label{tildeA1}
    \tilde{A}_{1}=A_{1}+\frac{i\omega A_1}{4r_{h}} \log(2r_{h})-\frac{i \pi \omega A_1}{8r_{h}}.
\end{equation}

Then, we will compute the constant $A_1$. In order to do this, let us first rewrite the solutions in regions \textbf{A} and \textbf{C} as:
\begin{align}
    h_{\omega}^{A}(r)&=A_{1}\frac{e^{\frac{-k}{2r^{2}}}}{r}e^{-i\omega r_{*}},\\
    h_{\omega}^{C}(r)&=A_{1}\left[\mathcal{C}_{1}+i\omega\left(\mathcal{C}_{2}+\frac{\mathcal{C}_{3}}{r^{3}}\right)\right],
\end{align}
where  
\begin{equation}\label{C1C2C3}
    \mathcal{C}_{1}=\frac{e^{-\frac{k}{2r_{h}^{2}}}}{r_{h}},\,\,\,
    \mathcal{C}_{2}=\frac{e^{-\frac{k}{2r_{h}^{2}}}}{r_{h}}\left(\frac{\log(2r_{h})+b}{4r_{h}}-\frac{\pi}{8r_{h}}\right), \,\, \,  \mathcal{C}_{3}=\frac{1}{3}e^{\frac{k}{2r_{h}^{2}}}r_{h}.
\end{equation}

The inner product between the solutions of Eq. \eqref{eq:EOM-NG} can be calculated by:
\begin{align}
    (X_{\omega}(r,t),X_{\omega}(r,t))&=\frac{-i}{2\pi \alpha'}\int_{r_{h}}^{r_{b}}dr\;\sqrt{\frac{g_{rr}}{-g_{tt}}}g_{xx} \left(h_{\omega}(r,t)\partial_{t}h_{\omega}^{*}(r,t)-(\partial_{t}h_{\omega}(r,t))h_{\omega}^{*}(r,t)\right)\nonumber \\
    &=\frac{\omega}{\pi \alpha'}\int_{r_{h}}^{r_{b}}dr\;\frac{e^{\frac{k}{r^{2}}}}{f(r)} |h_{\omega}(r)|^{2}=1. \label{IKGP}
\end{align}
In order to find an approximate solution for the above integral, note that the integrand is dominated by its behavior near the horizon where there is a logarithm divergence. Close to the horizon the blackening function, Eq. \eqref{HF}, is given by:
\begin{equation}
    f^{A}(r)=\left(1-\frac{r_{h}^{4}}{r^{4}}\right)=\frac{(r^{4}-r_{h}^{4})}{r^{4}}\approx \frac{4r_{h}^{3}}{r^{4}_{h}}(r-r_{h})=\frac{4(r-r_{h})}{r_{h}}.
\end{equation}
Then one gets:
\begin{align}
     \frac{\pi \alpha'}{\omega}\left[\frac{1}{4r_{h}}\int_{r_{h}+\epsilon}\frac{dr}{(r-r_{h})}\right]^{-1}&=|A_{1}|^{2},
\end{align}
where we disregarded the subleading term near the brane which depends explicitly on $r_b$. Performing the above integral, one obtains the normalization factor $A_{1}$:
\begin{align}\label{eq:A1final}
    A_{1}=\sqrt{\frac{4 \pi \alpha' r_{h}}{\omega  |\log{\epsilon}|}}=\sqrt{\frac{4 \pi \alpha' r_{h}}{\omega  \log\left(\frac{1}{\epsilon}\right)}}.
\end{align}
Then, the solution $h^C_\omega$ is finally written as 
\begin{equation}
    h_{\omega}^{C}(r)=\sqrt{\frac{4 \pi \alpha' r_{h}}{\omega  \log\left(\frac{1}{\epsilon}\right)}}\left[\mathcal{C}_{1}+i\omega\left(\mathcal{C}_{2}+\frac{\mathcal{C}_{3}}{r^{3}}\right)\right],\label{hComega}
\end{equation}
where $\mathcal{C}_1$, $\mathcal{C}_2$, and $\mathcal{C}_3$ are given by Eq. \eqref{C1C2C3}.

\section{Fluctuation-Dissipation theorem at $T\neq0$}\label{FDT}

\subsection{The linear response function}\label{amdtdif}

In this section we will compute the admittance $\chi(\omega)$. Let us consider a particle under the action of an external force in an arbitrary direction, $x^{i}$, given by
\begin{equation}\label{forcew}
F(t) = E e^{-i\omega t} F(\omega), 
\end{equation}
\noindent where $E$ is the electric field on the brane.  In order to deal with the electric field $E = E(A_t, \vec A)$ one has take it into account it in the  approximate Nambu-Goto action.  Explicitly, 
\begin{equation}\label{eq:actionBH+E}
S \approx - \frac{1}{4 \pi \alpha'} \int d\tau d\sigma \left[ \;\dot{X}^2 \frac{e^{\frac{ k}{r^2}}}{f(r)}-X'^2 r^4 f(r) e^{\frac{k}{r^2}} \right] + \int dt \left(A_t + \vec{A} \cdot \vec{\dot{x}} \right)\Big|_{r=r_b}.
\end{equation}
From the above equation one can see that second term, corresponding an electric energy density, is just a surface term, chosen in an arbitrary direction, and does not contribute to the bulk dynamics.

To compute the response function, we assume that the external force $F(t)$, given by Eq. \eqref{forcew}, is linearly  coupled to $X'(t,r)$ on the brane.  Rewriting the surface term in a convenient way we have
\begin{equation}\label{ngbtermBH+E}
S \approx - \frac{1}{4 \pi \alpha'} \int dt dr  \left[ \;\dot{X}^2 \frac{e^{\frac{ k}{r^2}}}{f(r)}-X'^2 r^4 f(r) e^{\frac{k}{r^2}} \right] - \int dt \; F(t)\,\left.\left(\frac{\partial X(t,r)}{\partial r}\right)\right|_{r=r_{b}}, 
\end{equation}
where we choose $\tau=t$ and $\sigma=r$. 
On the brane, the equation of motion, $\delta S/\delta X'=0$ implies 
\begin{equation}\label{force}
F(t) = \frac{1}{2 \pi \alpha'} \left[X'(t,r_b)\; (r_{b}^{4}-r_{h}^{4})  e^{\frac{ k}{r_b^2}}\right]. 
\end{equation}
 Hence, the Neumann boundary condition on the brane reads
\begin{equation}\label{eq:BCBH+E}
    X'(t,r_b)=\frac{2\pi \alpha'}{(r_{b}^{4}-r_{h}^{4})} e^{-\frac{k}{r_b^2}}F(t). 
\end{equation}
As we have chosen the ingoing boundary condition at $r=r_{h}$, we can find directly $X'(\omega,r_{b})$, using Eq. \eqref{hcfinal}
\begin{align}
    X'(\omega,r_{b})=\left.\frac{\partial h^{(C)}_{\omega}}{\partial r}\right|_{r=r_{b}}=-i\omega A_{1}\frac{r_{h}e^{\frac{k}{2r_{h}^{2}}}}{r_{b}^{4}}. 
\end{align}
So $F(\omega)$ reads
\begin{equation}
    F(\omega)= -\frac{i\omega A_{1}}{2 \pi \alpha'} \left[\frac{r_{h}}{r_{b}^{4}} (r_{b}^{4}-r_{h}^{4})  e^{\frac{k}{r_b^2}+\frac{k}{2r_h^2}}\right].
\end{equation}

\begin{figure}
	\centering
	\includegraphics[scale = 0.6]{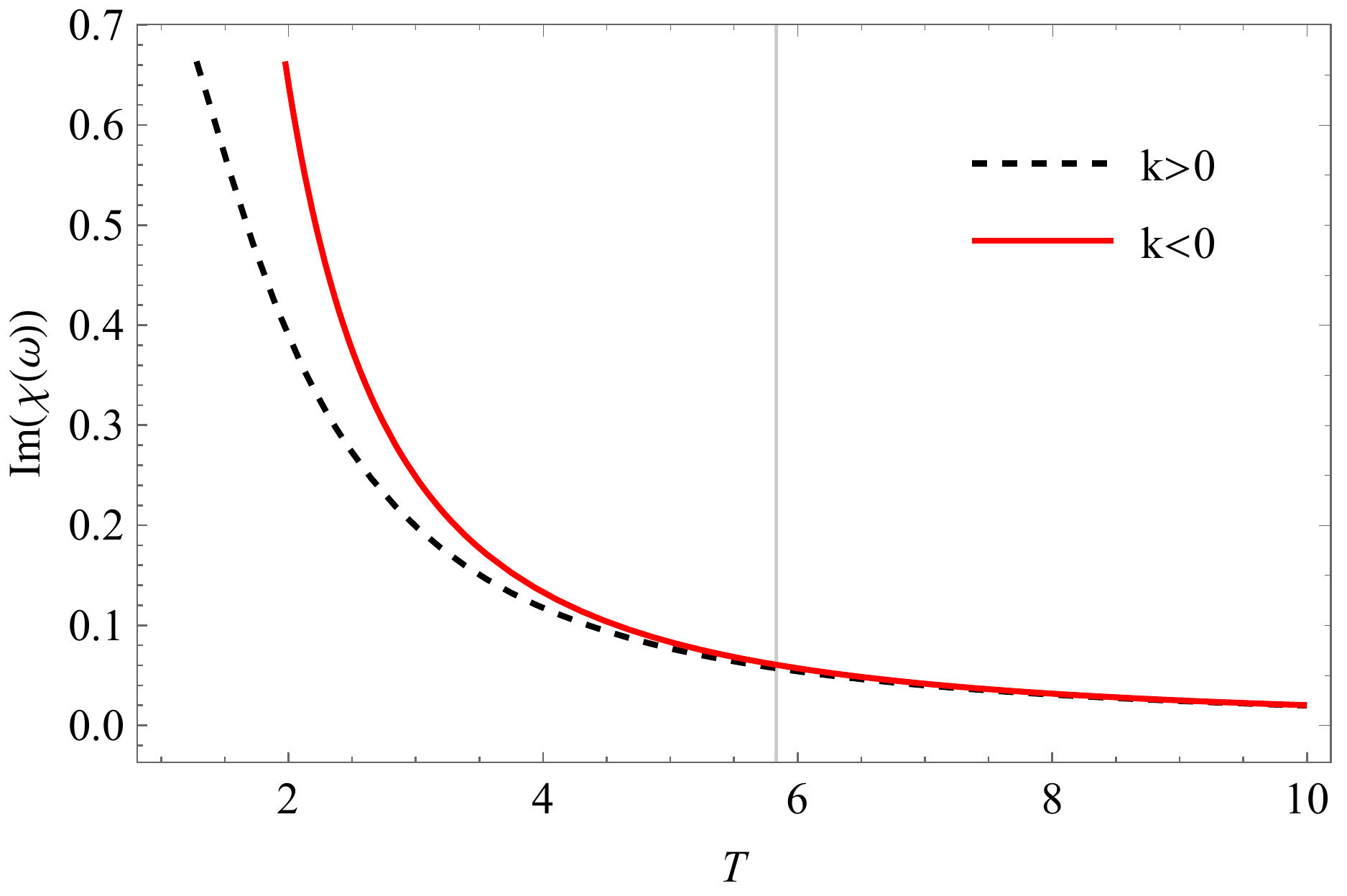}
	\caption{The imaginary part of admittance $\chi(\omega)$, for a fixed $\omega$, as a function of the  temperature for both $k = \pm 1$ in a arbitrary energy units from Eq. \eqref{eq:chitemperature}. The vertical line represents the approximate value for the temperature $ (T \approx 5.8)$. From this temperature forward (high temperatures) the sign of $k$ is no longer relevant.}
	\label{chi}
\end{figure}

In order to find the admittance, one notices that $\langle x(\omega)\rangle = h^{(C)}_{\omega}(r_{b})$, therefore
\begin{equation}
    \chi(\omega)=\frac{h^{(C)}_{\omega}(r_{b})}{F(\omega)}=\frac{\tilde{A}_{1} \frac{e^{-\frac{k}{2 r^2_h}}}{r_h} + \frac{i A_1  \omega}{4r_{h}^{2}}  e^{-\frac{k}{2 r_h^2}} b + \frac{ - \frac{1}{3} \left(-i A_1 r_{h} \omega  e^{\frac{k}{2 r^{2}_{h}}}\right)}{r^{3}_{b}}}{-\frac{i\omega A_{1}}{2 \pi \alpha'} \left[\frac{r_{h}}{r_{b}^{4}} (r_{b}^{4}-r_{h}^{4})  e^{\frac{k}{r_b^2}+\frac{k}{2r_h^2}}\right]}.
\end{equation}
Using the expression for $\tilde{A}_{1}$, Eq. \eqref{tildeA1}, one can expand $\chi(\omega)$ in the hydrodynamic limit $\omega\ll 1$ as
\begin{align}
    \chi(\omega)&=\frac{A_{1}\left\{\frac{e^{-\frac{k}{2 r^2_h}}}{r_h}+i\omega\left[\left(\frac{\log(2r_{h})+b}{4r_ {h}} -\frac{\pi}{8r_ {h}}\right) \frac{e^{-\frac{k}{2 r^2_h}}}{r_h} + \frac{  \left( r_{h}   e^{\frac{k}{2 r^{2}_{h}}}\right)}{3r^{3}_{b}}\right]\right\}}{-A_{1}\frac{i\omega}{2 \pi \alpha'} \left[\frac{r_{h}}{r_{b}^{4}} (r_{b}^{4}-r_{h}^{4})  e^{\frac{k}{r_b^2}+\frac{k}{2r_h^2}}\right]}\nonumber \\
    &\approx \frac{1}{-i\omega} \left[\frac{2\pi\alpha' e^{-k\left(\frac{1}{r_{b}^{2}}+\frac{1}{r_{h}^{2}}\right)}}{r_{h}^{2}f(r_{b})}\right] \xrightarrow{r_{b}\to\infty} \frac{1}{-i\omega} \left[\frac{2\pi\alpha' e^{-\frac{k}{r_{h}^{2}}}}{r_{h}^{2}}\right].\label{eq:chifinal}
\end{align}
By the definition of the temperature, Eq. \eqref{thor}, our admittance can be written as
\begin{equation}\label{eq:chitemperature}
    \chi(\omega)\approx \frac{1}{-i\omega} \left[\frac{2\alpha' e^{-k\frac{1}{r_{b}^{2}}}}{f(r_{b})}\right]\frac{e^{-\frac{k}{\pi^{2}T^{2}}}}{\pi T^{2}} \xrightarrow{r_{b}\to\infty} \frac{1}{-i\omega} \frac{2\alpha'}{\pi T^{2}}e^{-\frac{k}{\pi^{2}T^{2}}}.
\end{equation}
In order to recover the pure AdS case, one has to consider the limit $k\to0$. Then, we obtain that the AdS admittance is
\begin{equation}
    \chi(\omega)_{AdS} \approx \frac{1}{-i\omega} \frac{2\alpha'}{\pi T^{2}}, 
\end{equation}
in accordance with \cite{Giataganas:2018ekx}. 
 One can proceed the analysis of the admittance as  function of the sign of $k$. From Eq. (\ref{eq:chitemperature}), one finds that the ratio between the imaginary parts of the negative and positive signs of $k$ in the admittance  is given by
\begin{equation}
    \frac{\text{Im }\chi^{(k<0)}(\omega)}{\text{Im }\chi^{(k>0)}(\omega)}=e^{\frac{2|k|}{\pi^{2}T^{2}}}.
\end{equation}
Notice that the sign of $k$ is not important in the high temperature limit $T^2\gg |k|$. However, for the low temperature regime $T^2 \ll |k|$, the sign of $k$ is  relevant. This can be seen in Fig. \ref{chi} where the imaginary part of the admittance is plotted as a function of the temperature for the two different signs of $k$.

The diffusion constant can be obtained as
\begin{equation}\label{diffusion}
    D = T \lim_{\omega \to 0}(- i \omega \chi(\omega))=\left[\frac{2\alpha' e^{-k\frac{1}{r_{b}^{2}}}}{f(r_{b})}\right]\frac{e^{\frac{-k}{\pi^{2}T^{2}}}}{\pi T}\xrightarrow{r_{b}\to\infty} \frac{2\alpha' }{\pi T} e^{\frac{-k}{\pi^{2}T^{2}}}.
\end{equation}
Interestingly this result was obtained in  \cite{Dudal:2018rki} within a different model, where the a dilaton field is introduced directly in the Nambu-Goto action. Moreover, they obtained this result from the relation between the mean square displacement and the diffusion constant for the Brownian motion instead of the procedure performed here, where $D$ is obtained from the admittance. Indeed, in Sec. \ref{sec:meansquare} we also obtain the diffusion constant $D$ by this method. 

The AdS limit of the diffusion constant reads
\begin{equation}
D_{AdS}=\lim_{k\to0} \frac{2\alpha' e^{\frac{-k}{\pi^{2}T^{2}}}}{\pi T} = \frac{2\alpha'}{\pi T}.
\end{equation}
This is the diffusion constant for the AdS with $T\neq0$ already obtained in Refs. \cite{deBoer:2008gu,Dudal:2018rki}.

Following Ref. \cite{Giataganas:2018ekx}, it is interesting to expand $\chi(\omega)$ up to order $\omega$. From Eq. \eqref{eq:chifinal}, we find
\begin{equation}\label{eq:chicompleto}
    \chi(\omega)=\frac{2\pi\alpha' }{-i\omega \left[r_{h}^{2} f(r_{b})  e^{\frac{k}{r_b^2}+\frac{k}{r_h^2}}\right]}-\frac{2\pi\alpha'\left[\left(\frac{\log(2r_{h})+b}{4r_ {h}} -\frac{\pi}{8r_ {h}}\right) \frac{e^{-\frac{k}{2 r^2_h}}}{r_h} + \frac{  \left( r_{h}   e^{\frac{k}{2 r^{2}_{h}}}\right)}{3r^{3}_{b}}\right]}{\left[r_{h} f(r_{b}) e^{\frac{k}{r_b^2}+\frac{k}{2r_h^2}}\right]}+\mathcal{O}(\omega).
\end{equation}

Notice that in Ref. \cite{Giataganas:2018ekx}, it was proposed that the admittance in the low frequency expansion limit and $r_{b}\to\infty$, in a general metric, can be written as
\begin{equation}
    \chi(\omega)=\frac{2\pi\alpha'}{-i\omega g_{xx}(r_{h})}.
\end{equation}
Indeed this expression is recovered by our result Eq. \eqref{eq:chicompleto} where we identify $g_{xx}(r_{h})=r_{h}^{2}\exp({k/r_{h}^{2}})$.

Further, comparing Eq. \eqref{eq:chicompleto} to the general expansion of $\chi(\omega)$ presented in \cite{Giataganas:2018ekx}
\begin{equation}
    \chi(\omega)=2\pi\alpha'\left(\frac{i}{\gamma \omega} - \frac{m}{\gamma^{2}} +\mathcal{O}(\omega)\right),
\end{equation}
one finds that the self-energy of the particle is 

\begin{equation}{\label{eq:gamma}}
    \gamma = r_{h}^{2} f(r_{b}) e^{\frac{k}{r_{b}^{2}}+\frac{k}{r_{h}^{2}}}= \pi^{2} T^{2} e^{\frac{k}{r_{b}^{2}}}f(r_{b},T)e^{\frac{k}{\pi^{2}T^{2}}} \xrightarrow{r_{b}\to \infty} \pi^{2} T^{2} e^{\frac{k}{\pi^{2}T^{2}}}.
\end{equation}

The inertial mass reads:
\begin{align}
    m&=\left[\left(\frac{\log(2r_{h})+b}{4r_ {h}} -\frac{\pi}{8r_ {h}}\right) \frac{e^{-\frac{k}{2 r^2_h}}}{r_h} + \frac{  \left( r_{h}   e^{\frac{k}{2 r^{2}_{h}}}\right)}{3r^{3}_{b}}\right] \left(r_{h}^{3} f(r_{b}) e^{\frac{k}{r_{b}^{2}}+\frac{3k}{2r_{h}^{2}}} \right)\nonumber \\
    &\quad \xrightarrow{r_{b}\to\infty} \left(\frac{\log(2r_{h})+b}{4} -\frac{\pi}{8}\right) r_{h} f(r_{b}) e^{\frac{k}{r_{b}^{2}}+\frac{k}{r_{h}^{2}}} \nonumber \\
    &= \left(\frac{\log(2\pi T)+b}{4} -\frac{\pi}{8}\right) \pi T e^{\frac{k}{\pi^{2}T^{2}}}.
\end{align}

To conclude this subsection it is interesting to compare our results with Refs. \cite{Tong:2012nf,Edalati:2012tc,Giataganas:2018ekx}. As can be seen in $h_\omega^C$, Eq. \eqref{hComega},  in the  admittance, Eq. \eqref{eq:chitemperature} and in the transport coefficient $D$, Eq. \eqref{diffusion}, these quantities  can not be obtained from  a polynomial metric as in Refs. \cite{Tong:2012nf,Edalati:2012tc,Giataganas:2018ekx}. However, in the asymptotic limit they are related by a regular exponential factor  $e^{\frac{k}{r_h^{2}}}$.

\subsection{Thermal two-point function for the string end-point at the brane} \label{displace}

In this subsection the thermal two-point function for the end-point of the string located at the brane will be obtained. 
By using a Fourier decomposition, such as:
\begin{equation}\label{fourrier}
X(t,r) = \int_0^{\infty }d\omega \,  \left(h^C_{\omega }(r)e^{-\text{i$\omega t $}} a_{\omega} + h^{C*}_{\omega}(r)e^{\text{i$\omega t $}}   a^{\dagger}_{\omega}\right) \, ,
\end{equation}
where $a_{\omega}$ and $a^{\dagger}_{\omega}$ are the annihilation and creation operators, respectively. Recalling that, for $T\neq0$, one has:
\begin{eqnarray}
\label{ValorEsperado}
\langle a^{\dagger}_{\omega}a_{\omega} \rangle
&=& \text{Tr}\,(e^{-\beta \sum \omega_n}a^{\dagger}_{\omega}a_{\omega'})=\frac{\delta_{\omega\omega'}}{e^{\beta\omega}-1}, \nonumber \\
\langle a^{\dagger}_{\omega_n}a^{\dagger}_{\omega} \rangle
&=&\text{Tr}\,(e^{-\beta \sum \omega_n}a^{\dagger}_{\omega}a^{\dagger}_{\omega'})=0, \,\, \langle a_{\omega}a_{\omega} \rangle
= \text{Tr}\,(e^{-\beta \sum \omega_n}a_{\omega}a_{\omega'})=0,
\end{eqnarray}
\noindent which represent the expected values of the product between the creation and annihilation operators with a Bose-Einstein factor. Identifying $x(t)=X(t,r_{b})$, one gets:
\begin{align}
\langle x(t)x(0) \rangle&=\langle X(t,r_{b})X(0,r_{b}) \rangle\nonumber \\
&=\left\langle\sum_{\omega>0}\sum_{\omega'>0}\left(h^{C}_{\omega}(r_b)e^{-i\omega t}a_{\omega}+h^{C*}_{\omega}(r_b)e^{i\omega t}a^{\dagger}_{\omega} \right)\left(h^{C}_{\omega'}(r_b)a_{\omega'}+h^{C*}_{\omega'}(r_b)a^{\dagger}_{\omega'} \right)\right\rangle\nonumber \\
&= \sum_{\omega>0}
|h^{C}_{\omega}(r_b)|^{2}
\Big( \frac{2\cos(\omega t)}{e^{\beta\omega}-1}
+ e^{-i\omega  t}\Big)\nonumber \\&=\frac{4 \pi r_{h} \alpha'}{\log\left(\frac{1}{\epsilon}\right)}\sum_{\omega>0}\frac{1}{\omega}
\left(\mathcal{C}_{1}^{2}+\omega^{2}\mathcal{C}_{2}^{2}\right)
\Big( \frac{2\cos(\omega t)}{e^{\beta\omega}-1}
+ e^{-i\omega  t}\Big),  \label{eq:CorrelationPosition}
\end{align}
where we used the solution $h^C_\omega(r)$ given by Eq. \eqref{hComega}. 
Using Eq. \eqref{eq:discretenesscondition}, this discrete sum can be approximated by an integral 
\begin{equation}
\sum_{\omega>0} \Delta \omega \longrightarrow \int_{0}^{\infty}d\omega \Leftrightarrow \sum_{\omega>0} \frac{4\pi r_{h}}{\log\left(\frac{1}{\epsilon}\right)} \longrightarrow \int_{0}^{\infty}d\omega. 
\end{equation}
Therefore the correlation function at the brane reads
\begin{equation}\label{correlxtx0}
 \langle x(t)x(0) \rangle =   \alpha' \int_{0}^{\infty}\frac{d\omega}{\omega}
\left(\mathcal{C}_{1}^{2}+\omega^{2}\mathcal{C}_{2}^{2}\right)
\Big( \frac{2\cos(\omega t)}{e^{\beta\omega}-1}
+ e^{-i\omega  t}\Big).
\end{equation}
This is the thermal two-point function for the string endpoint at the brane. 

\subsection{The mean square displacement}\label{sec:meansquare}

From the thermal two-point function  for the endpoint of the string at the brane, Eq. \eqref{correlxtx0}, one can compute the mean square displacement:
\begin{align}
\label{MeanSquareTerms}
s^{2}(t)&\equiv\langle[x(t)-x(0)]^{2}\rangle=\langle x(t)^{2}\rangle + \langle x(0)^{2}\rangle - \langle x(t)x(0)+x(0)x(t)\rangle.
\end{align}
Each term will be computed separately
\begin{align}\label{eq:xmeansquaret}
    \langle x(t)^{2} \rangle &= \sum_{\omega>0}\sum_{\omega'>0}\left\langle\left[\left(h^{C}_{\omega}(r_{b})e^{-i\omega t}a_{\omega}+h^{C*}_{\omega}(r_{b})e^{i\omega t}a^{\dagger}_{\omega} \right)\left(h^{C}_{\omega'}(r_{b})e^{-i\omega' t}a_{\omega'}+h^{C*}_{\omega'}(r_{b})e^{i\omega' t}a^{\dagger}_{\omega'} \right)\right]\right\rangle \nonumber \\
              &=\frac{4\pi r_{h} \alpha'}{\log\left(\frac{1}{\epsilon}\right)}\sum_{\omega>0} \frac{1}{\omega} \left(\mathcal{C}_{1}^{2}+\omega^{2}\mathcal{C}_{2}^{2}\right)\left(\frac{2}{e^{\beta \omega} -1}+1\right)\nonumber \\ &=\alpha' \int_{0}^{\infty}\frac{d\omega}{\omega}\left(\mathcal{C}_{1}^{2}+\omega^{2}\mathcal{C}_{2}^{2}\right)\left(\frac{2}{e^{\beta \omega} -1}+1\right).
\end{align}
By the same token one finds
\begin{equation}\label{eq:xmeansquare0}
      \langle x(0)^{2} \rangle =\alpha' \int_{0}^{\infty}\frac{d\omega}{\omega}\left(\mathcal{C}_{1}^{2}+\omega^{2}\mathcal{C}_{2}^{2}\right)\left(\frac{2}{e^{\beta \omega} -1}+1\right)= \langle x(t)^{2} \rangle.
\end{equation}
We have already computed $\langle x(t)x(0)\rangle$ in Eq. \eqref{eq:CorrelationPosition}. The last two-point correlation function is
    \begin{align}
\label{eq:antiCorrelationPosition}
\langle x(0)x(t) \rangle&=\left\langle\sum_{\omega>0}\sum_{\omega'>0}A_{1\omega}A_{1\omega'}\left(h^{C}_{\omega}(r_{b})a_{\omega}+h^{C*}_{\omega}(r_{b})a^{\dagger}_{\omega} \right)\left(h^{C}_{\omega'}(r_{b})e^{-i\omega' t}a_{\omega'}+h^{C*}_{\omega'}(r_{b})e^{+i\omega' t}a^{\dagger}_{\omega'} \right)\right\rangle\nonumber \\
&=\frac{4 \pi r_{h} \alpha'}{ \log\left(\frac{1}{\epsilon}\right)}\sum_{\omega>0}\frac{1}{\omega}
\left(\mathcal{C}_{1}^{2}+\omega^{2}\mathcal{C}_{2}^{2}\right)
\Big( \frac{2\cos(\omega t)}{e^{\beta\omega}-1}
+ e^{i\omega  t}\Big) \nonumber \\
&=\alpha'\int_{0}^{\infty}\frac{d\omega}{\omega}
\left(\mathcal{C}_{1}^{2}+\omega^{2}\mathcal{C}_{2}^{2}\right)
\Big( \frac{2\cos(\omega t)}{e^{\beta\omega}-1}
+ e^{i\omega  t}\Big).
\end{align}
Collecting together these results one  obtains
\begin{align}
    s^{2}(t)&=\langle x(t)^{2}\rangle + \langle x(0)^{2}\rangle - \langle x(t)x(0)\rangle- \langle x(0)x(t)\rangle\nonumber \\
    &=\alpha'\int_{0}^{\infty}\frac{d\omega}{\omega}
\left(\mathcal{C}_{1}^{2}+\omega^{2}\mathcal{C}_{2}^{2}\right)\Big[ \frac{4(1-\cos(\omega t))}{e^{\beta\omega}-1}+(2-e^{i\omega t}-e^{-i\omega t})\Big]\nonumber \\
&=\alpha'\int_{0}^{\infty}\frac{d\omega}{\omega}\left(\mathcal{C}_{1}^{2}+\omega^{2}\mathcal{C}_{2}^{2}\right) \coth\left(\frac{\beta \omega}{2}\right) \sin^{2}\left(\frac{\omega t}{2}\right). \label{eq:s^2naoren}
\end{align}
\noindent 
This expression for the mean square displacement diverges. Hence, by using the normal ordering  one can write a regularized mean square displacement as
\begin{equation}
    s^2_{\rm reg}(t) = \langle : [x(t) - x(0)]^2 : \rangle = \langle : [X(t, r_b) - X(0, r_b)]^2 : \rangle. 
\end{equation}{}
%%%%%%%%%%%%%%%%%%%%%%%%%%%%%%
Note that, in the normal ordering, one has $\langle a^{\dagger}_{\omega}a_{\omega'} \rangle=\langle a_{\omega}a^{\dagger}_{\omega'} \rangle=\delta_{\omega\omega'}(e^{\beta\omega}-1)^{-1}$.

%%%%%%%%%%%%%%%

Repeating the steps performed to obtain  Eq.  \eqref{eq:s^2naoren}, the regularized mean square displacement is obtained:
\begin{equation}\label{disp2}
s^2_{\rm reg}(t) = \alpha'\int_{0}^{\infty}\frac{d\omega\, \mathcal{C}_1^2}{\omega}\left[\frac{2}{e^{\beta \omega}- 1} \right]  \sin^{2}\left(\frac{\omega t}{2}\right) + \alpha'\int_{0}^{\infty}d\omega\, \mathcal{C}_2^2\, \omega\left[\frac{2}{e^{\beta \omega}- 1} \right]  \sin^{2}\left(\frac{\omega t}{2}\right), 
\end{equation}
\noindent or in a more compact way 
\begin{equation}
s^2_{\rm reg}(t) = \alpha' \left[ {\cal I}_1(t) + {\cal I}_2(t)\right],
\end{equation}
where
\begin{align}
\mathcal{I}_{1}&=\int_{0}^{\infty}\frac{d\omega\, \mathcal{C}_1^2}{\omega}\left[\frac{2}{e^{\beta \omega}- 1} \right]  \sin^{2}\left(\frac{\omega t}{2}\right), \label{eq:I1} \\
\mathcal{I}_{2}&=\int_{0}^{\infty}d\omega\, \mathcal{C}_2^2\, \omega\left[\frac{2}{e^{\beta \omega}- 1} \right]  \sin^{2}\left(\frac{\omega t}{2}\right).\label{eq:I2}
\end{align}
 The integral \eqref{eq:I1} can be cast in the form
\begin{equation}
  {\cal I}_1(t) =  2 \alpha' \mathcal{C}_1^2 \sum_{n = 1}^{\infty} \int_{0}^{\infty}\frac{d\omega}{\omega}e^{-\beta \omega n}\sin^{2}\left(\frac{\omega t}{2}\right)\,, \label{eq:I1_part2}
\end{equation}{}
where we have used the following identity 
\begin{equation}\label{eq:geometricseries}
   \frac{1}{e^{\beta \omega}- 1}  =  \frac{e^{-\beta \omega}}{1- e^{-\beta \omega}} = \sum_{n = 0}^{\infty}e^{-\beta \omega (n+1)}\,.
\end{equation}{}
Equation \eqref{eq:I1_part2} can be integrated:
\begin{equation}
    {\cal I}_1(t) = \frac{2 \alpha' \mathcal{C}_1^2}{4} \left[\sum_{n = 1}^{\infty} \log \left(1 + \frac{t^2}{n^2 \beta^2} \right)\right]= \frac{ \alpha' \mathcal{C}_1^2}{2} \log \left[\prod_{n=1}^{\infty}\left(1 + \frac{t^2}{ n^2 \beta^2} \right)\right].
\end{equation}{}
Using the identity
\begin{equation}
    \frac{\sinh z}{z} = \prod_{n=1}^{\infty}\left(1 + \frac{z^2}{\pi^2 n^2 } \right)\,
\end{equation}{}
\noindent one gets:
\begin{equation}\label{I1}
     {\cal I}_1(t) = \frac{ \alpha' \mathcal{C}_1^2}{2}  \log \left(\frac{\sinh (\frac{t \pi}{\beta})}{\frac{t \pi}{\beta}} \right).
\end{equation}{}

Now we have to deal with the second integral, $\mathcal{I}_{2}$, in Eq. \eqref{eq:I2}, by using the identity \eqref{eq:geometricseries}, one finds:
\begin{eqnarray}
     {\cal I}_2 &=& 2 \alpha' \mathcal{C}^2_2 \int_{0}^{\infty}d\omega\, \omega \left[\frac{\sin^{2}\left(\frac{\omega t}{2}\right)}{e^{\beta \omega}- 1} \right]   =  2 \alpha' \mathcal{C}^2_2 \sum_{n=1}^{\infty}\int_{0}^{\infty}d\omega \,\omega\, e^{\beta \omega n}\sin^{2}\left(\frac{\omega t}{2}\right) \nonumber \\
     &=& 2 \alpha' \mathcal{C}^2_2 \sum_{n=1}^{\infty} \frac{t^4 + 3 n^2 t^2 \beta^2}{2 n^2 \beta^2 (t^2 + n^2 \beta^2)^{2}}. \label{I2}
\end{eqnarray}{}

Now, one can investigate whether our deformed string/gauge setup has a ballistic as well as diffusive regimes. 
Then, one has to consider the  appropriate limits for very short and long times.

From equation \eqref{I1} one can analyze the short time limit, $t \ll \beta/\pi$,  for ${\cal I}_1(t)$: 
\begin{equation}
    \sinh \left( \frac{t \pi}{\beta}\right)  \approx \frac{t \pi}{\beta} + \frac{t^3 \pi^3}{3! \beta^3}\,,
\end{equation}{}
\noindent then 
\begin{equation}\label{I1B}
     {\cal I}_1(t) \approx \frac{ \alpha' \mathcal{C}_1^2}{2}  \log \left(1 +  \frac{t^2 \pi^2}{3! \beta^2} \right)\approx \frac{ \alpha' \pi^2 \mathcal{C}_1^2 }{12 \beta^2} t^2\,.
\end{equation}{}
For the long time limit, $t \gg {\beta}/{\pi}$, the  expression  \eqref{I1} can be approximated by
\begin{eqnarray}
    \log \left(\frac{\sinh (\frac{t \pi}{\beta})}{\frac{t \pi}{\beta}} \right) &=& \log \left(\sinh \left(\frac{t \pi}{\beta}\right)\right) - \log \left(\frac{t \pi}{\beta}\right) \approx \frac{t \pi}{\beta} - \log \left(\frac{t \pi}{\beta}\right)
    \approx  \frac{t \pi}{\beta}
\end{eqnarray}{}
Therefore in this limit, one obtains
\begin{equation}
    {\cal I}_1(t) \approx \frac{ \alpha' \pi \mathcal{C}_1^2 }{2 \beta} t.
\end{equation}

For ${\cal I}_2(t)$, one can analyze the regimes $t \ll \beta/\pi$ and  $t \gg \beta/\pi$. For the short time limit, Eq. \eqref{I2} becomes 
\begin{eqnarray}
    {\cal I}_2 &=& 2 \alpha' \mathcal{C}^2_2 \sum_{n=1}^{\infty} \frac{3 t^2 }{2 n^4 \beta^4}= \frac{3 \alpha' \mathcal{C}^2_2}{\beta^4} t^2 \sum_{n=1}^{\infty} \frac{1}{n^4 }= \frac{3 \alpha' \mathcal{C}^2_2 \zeta(4)}{\beta^4} t^2,
\end{eqnarray}{}
\noindent where $\zeta(s)$ is the Riemann zeta function \cite{abramowitz+stegun}. On the other side, for the long time limit, $t \gg {\beta}/{\pi}$, Eq. \eqref{I2} reads
\begin{eqnarray}
    {\cal I}_2 \approx  2 \alpha' \mathcal{C}^2_2 \sum_{n=1}^{\infty} \frac{1}{2 n^2 \beta^2} = {\rm const.}
\end{eqnarray}

The importance of those limits, $t \ll {\beta}/{\pi}$ and $t \gg {\beta}/{\pi}$, relies upon that for the Brownian motion the short time limit represents the ballistic regime and the long time limit represents the diffusive one. First to study the ballistic regime one has to take into account the contribution from  ${\cal I}_1$ and $ {\cal I}_2 $ for $t \ll {\beta}/{\pi}$.
\begin{eqnarray}
    s^2_{\rm reg}(t)& =& {\cal I}_1 +  {\cal I}_2 =  \frac{ \alpha' \pi^2 \mathcal{C}_1^2 }{12 \beta^2} t^2 + \frac{3 \alpha' \mathcal{C}^2_2 \zeta(4)}{\beta^4} t^2 \label{bali}\,,
\end{eqnarray}{}
\noindent where $\zeta(4) = \pi^4/90$, $\mathcal{C}_1$ and $\mathcal{C}_2$ are given by Eq. \eqref{C1C2C3} and $\beta = 1/T =\pi/r_h$.  
Then one can write Eq. \eqref{bali} for ballistic regime as :
\begin{eqnarray}\label{balfinal}
    s^2_{\rm reg}(t)& =& \frac{ \alpha' e^{\frac{-k }{T^2 \pi^2}}  }{6}\left[ \frac{1}{2} + \frac{1}{80}\left(\log(2 \pi T) - \frac{\pi}{2}\right)^2\right]t^2.
\end{eqnarray}{}
Notice that, for the short time limit, one recovers the ballistic regime, $s^2_{\rm reg}(t)\sim t^{2}$. On the other hand, the long time limit is given by the contribution from  ${\cal I}_1$ and $ {\cal I}_2 $ for $t \gg {\beta}/{\pi}$. But in this regime only the contribution coming from ${\cal I}_1$ is relevant, so that:
\begin{equation}\label{diffinal}
    s^2_{\rm reg}(t) = {\cal I}_1  =  \frac{\alpha' \pi \mathcal{C}_1^2 }{2 \beta} t = \frac{ \alpha' e^{\frac{- k}{\pi^2 T^2}}}{2  \pi T} t \sim D t.
\end{equation}
Then, we recovered the diffusive regime, $s^2_{\rm reg}(t)\sim D t$, where $D$ is the diffusion constant given by Eq. \eqref{diffusion}.  Therefore, in this deformed string/gauge setup, we find the expected ballistic and diffusive regimes for the Brownian motion.

\subsection{Fluctuation-dissipation theorem}\label{teofd}

In our setup, one can check explicitly the fluctuation-dissipation theorem. In Fourier variables, this theorem can be stated as
\begin{equation}
    \langle x(\omega)x(0) \rangle = (2 n_{B}(\omega) + 1) {\rm Im}\,\chi(\omega).
\end{equation}
\noindent where ${\rm n_B}(\omega) = (e^{\beta \omega}-1)^{-1}$ is the Bose-Einstein distribution, related to thermal noise effects. Then one gets:
\begin{equation}
    \langle x(t)x(0) \rangle = \frac{1}{2\pi}\int_{0}^{\infty} d\omega \langle x(\omega)x(0)\rangle e^{-i\omega t}.
\end{equation}
Comparing the above equation with Eq. \eqref{correlxtx0}, one gets for small frequencies
\begin{align}\label{fludis}
    \langle x(\omega)x(0)\rangle &= \frac{2\pi \alpha' \mathcal{C}_{1}^{2}}{\omega}(2n_{B}(\omega)+1) = \underbrace{\frac{2\pi \alpha' e^{-\frac{k}{r_{h}^{2}}}}{\omega r_{h}^{2}}}_{\text{Im}\;\chi(\omega)}(2n_{B}(\omega)+1).
\end{align}

From the imaginary part of the admittance, Eq. \eqref{eq:chifinal},  we therefore have  verified the fluctuation-dissipation theorem in our setup.  
This result could be expected within our conformally deformed theory (asymptotically AdS) as also captured with the polynomial metric of Ref. \cite{Giataganas:2018ekx}. 

Finally, note that in the finite temperature scenario our results are smooth in the limit $k\to 0$ recovering the pure AdS case.

\section{Zero Temperature Scenario}\label{Tzero}

In this section we will present the linear response function at zero temperature. 
In this case the metric is given by
\begin{equation}\label{adst0}
ds^{2}= e^{\frac{k}{r^{2}}} \, r^{2} (\eta_{\mu\nu} dx^{\mu}dx^{\nu}+\frac{dr^{2}}{r^{4}}), 
\end{equation}
and Regge-Wheeler radial coordinate $r_*$ can be defined by 
\begin{equation}
    dr^{2}_{*}=\frac{dr^{2}}{r^{4}} \Rightarrow \frac{dr_{*}}{dr}=\pm\frac{1}{r^{2}} \Rightarrow r_{*}=\mp \frac{1}{r},
\end{equation}
where we disregarded an integration constant and we choose the  positive sign, $r_{*}=r^{-1}$. Now the region $r\sim 0$ is mapped to $r_{*}\sim \infty$ while $r\to \infty$ is identified with $r\to 0$. This $r_{*}$ coordinate is equivalent to the $z$ coordinate of the Poincaré patch which is extensively used in the context of AdS/CFT correspondence.

Using this coordinate, the line element is
\begin{equation}
    ds^{2}=
    \frac{e^{kr^{2}_{*}}}{r_*^2}\left(\eta_{\mu \nu} dx^{\mu}dx^{\nu}+dr_{*}^{2}\right).
\end{equation}
Thus, the equation of motion in Fourier space analogous to Eq. \eqref{ansatz} is
\begin{equation}
\label{PoincareMotionEquation}
    \frac{e^{kr_{*}^{2}}}{r_{*}^{2}}\omega^{2}h_{\omega}(r_{*})+\frac{d}{d r_{*}}\left(\frac{e^{kr_{*}^{2}}}{r_{*}^{2}}\frac{dh_{\omega}(r_{*})}{d r_{*}}\right)=0.
\end{equation}

Here we are interested in the low frequency regime then we will expand the solution of this equation in powers of the frequency $\omega$ (hydrodynamic expansion), so we can write
\begin{equation}
    h_{\omega}(r_{*})=h_{0}(r_{*})+\omega h_{1}(r_{*})+\mathcal{O}(\omega^{2}).
\end{equation}
Substituting the above equation into Eq. \eqref{PoincareMotionEquation}, one finds at each order 
\begin{align}
    \frac{d}{dr_{*}}\left(\frac{e^{kr_{*}^{2}}}{r_{*}^{2}}\frac{dh_{0}(r_{*})}{dr_{*}}\right)&=0, \label{h0}\\
    \frac{d}{dr_{*}}\left(\frac{e^{kr_{*}^{2}}}{r_{*}^{2}}\frac{dh_{1}(r_{*})}{dr_{*}}\right)&=0. \label{h1}
\end{align}
These equations can be solved promptly but separately for the cases $k>0$ and $k<0$. 

\subsection{The case $k<0$}

In the case $k<0$, we can solve Eqs. \eqref{h0} and \eqref{h1} to find 

\begin{eqnarray}
\label{HydrodynamicExpansiont0i}
    h_{0}(r_{*})=C_{1}+C_{0}\left(\frac{r_{*} e^{r_{*}^2 \left| k\right| }}{2 \left| k\right| }-\frac{\sqrt{\pi } \text{erfi}\left(r_{*} \sqrt{\left| k\right| }\right)}{4 \left| k\right| ^{3/2}}\right),\nonumber\\
    h_{1}(r_{*})=C^{(1)}_{1}+C^{(1)}_{0}\left(\frac{r_{*} e^{r_{*}^2 \left| k\right| }}{2 \left| k\right| }-\frac{\sqrt{\pi } \text{erfi}\left(r_{*} \sqrt{\left| k\right| }\right)}{4 \left| k\right| ^{3/2}}\right),
\end{eqnarray}
where $C_{0}$, $C_{1}$,  $C^{(1)}_{0}$ and $C^{(1)}_{1}$ are  independent of $\omega$ and $r_{*}$.

Therefore the solution for Eq. \eqref{PoincareMotionEquation} up the second order in $\omega$ is 
\begin{align}
\label{HHydrodynamicExpansiont0}
    h_{\omega}(r_{*})=&C_{1}+C_{0}\left(\frac{r_{*} e^{r_{*}^2 \left| k\right| }}{2 \left| k\right| }-\frac{\sqrt{\pi } \text{erfi}\left(r_{*} \sqrt{\left| k\right| }\right)}{4 \left| k\right| ^{3/2}}\right)\nonumber\\
    &+\omega\left(C^{(1)}_{1}+C^{(1)}_{0}\left(\frac{r_{*} e^{r_{*}^2 \left| k\right| }}{2 \left| k\right| }-\frac{\sqrt{\pi } \text{erfi}\left(r_{*} \sqrt{\left| k\right| }\right)}{4 \left| k\right| ^{3/2}}\right)\right)+\mathcal{O}(\omega^{2}).
\end{align}

Using the Bogoliubov  transformation
\begin{equation}\label{eq:Bogoliubovk}
    h_{\omega}(r_{*})=e^{B(r_{*})}\psi(r_{*})=\frac{e^{-\frac{kr_{*}^{2}}{2}}}{r_{*}}\psi(r_{*}),
\end{equation}
the $\psi(r_{*})$ part of the mode will satisfy the Schr\"odinger equation
\begin{equation}\label{eq:Schrodinger}
    \frac{d^{2}\psi(r_{*})}{d r_{*}^{2}}+(\omega^{2} - V(r_*))\psi(r_{*})=0, 
\end{equation} 
 with the potential
\begin{equation}\label{eq:potencialr*}
    V(r_{*})= -k +\frac{2}{r_{*}^{2}}+k^{2}r_{*}^{2}.
\end{equation}

\begin{figure}[h!]
    \centering
    \includegraphics[scale=0.75]{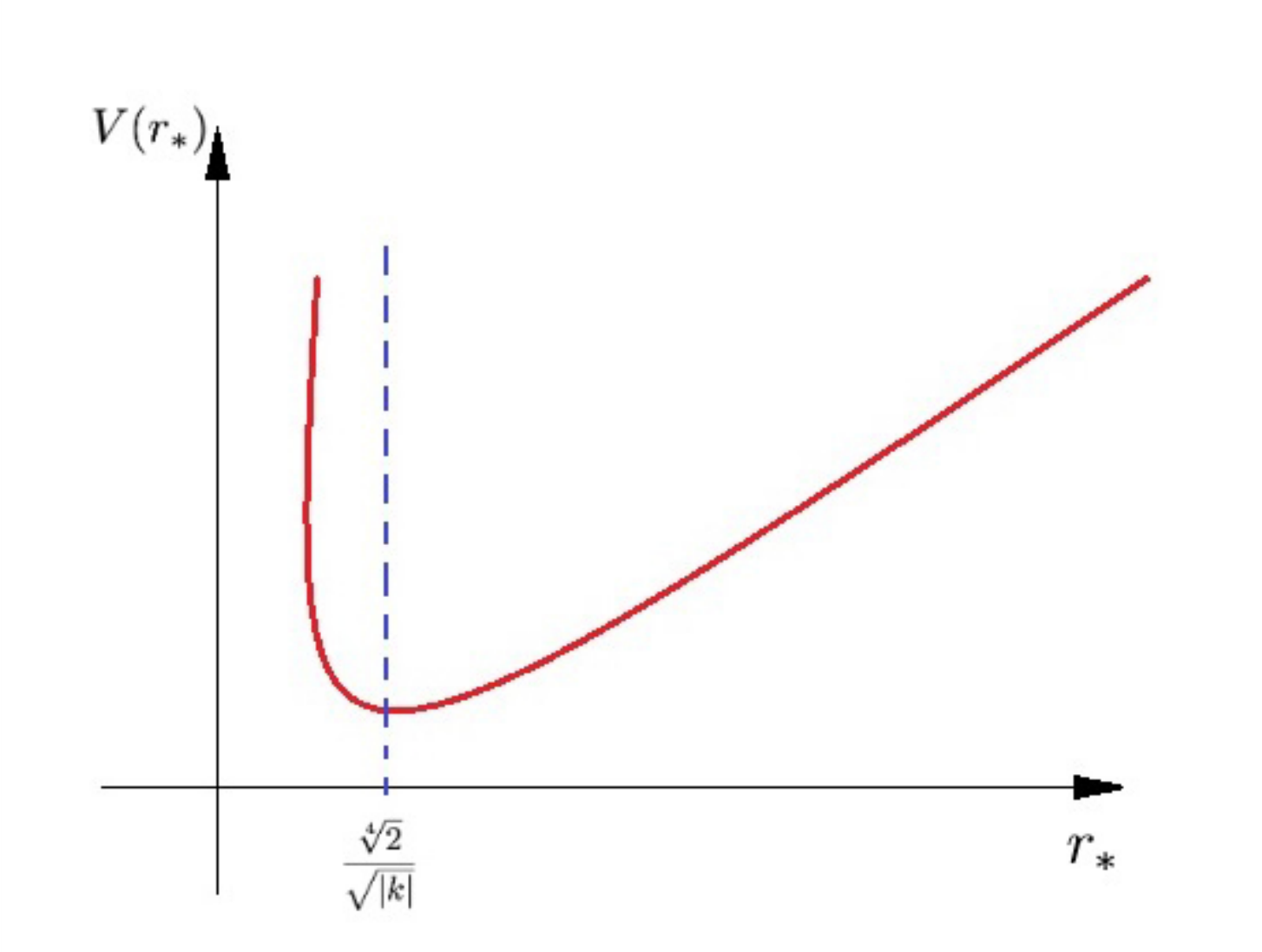}
    \caption{Sketch of the Potential $V(r_{*})$. Notice that this sketch is valid for both signals of $k$.}
    \label{fig:SketchV(r)}
\end{figure}

This potential has a minimum at $r_{*}=  r_{*\rm{min}}$, as sketched in Fig. \ref{fig:SketchV(r)}, where 
\begin{equation}
    r_{*\, \rm  min}=\frac{\sqrt[4]{2}}{\sqrt{|k|}},  
\end{equation}
and its value for $k<0$ is given by
\begin{align}
    V_{\rm min}=\left(2 \sqrt{2}+1\right) \left| k\right|. 
\end{align}

Since we are interested in the hydrodynamic limit of small $\omega$ we will consider the approximation $V(r_*)\sim V_{\rm min}$. 
Then,  the Schr\"odinger  equation  \eqref{eq:Schrodinger} in this limit becomes 
\begin{equation}\label{eq:SchrodingerCloseMinimum}
    \frac{d^{2}\psi(r_{*})}{d r_{*}^{2}}+(\omega^{2} - V_{\rm min})\psi(r_{*})=0. 
\end{equation} 
Therefore in the vicinity of $r_{*}\sim   r_{*\rm{min}}$, the solution is
\begin{equation}
    \psi(r_{*})=A_{1}e^{i r_{*}  \sqrt{\omega^{2}-V_{\rm min}}\, }+A_{2}e^{-i  r_{*} \sqrt{\omega^{2}-V_{\rm min}}\,}.
\end{equation}

Here we are going to work in the approximation $\omega^{2}\gg V_{\rm min}$. That approximation is good if $|k|/\omega^{2}\ll 1$ therefore for energies bigger than $\sqrt{|k|}$. This is expected since the value $\sqrt{|k|}$ can be seen as the natural energy scale of our setup. 
Thus $\psi(r_{*})$ can be written as
\begin{equation}
    \psi(r_{*})=A_{1}e^{i\omega r_{*}} +A_{2}e^{-i\omega r_{*}}. 
\end{equation}
 
Now we can write the general expression for $h_{\omega}(r_{*})$, Eq. \eqref{eq:Bogoliubovk}, as the solution close to the minimum $r_{*\, \rm min}$ of the potential, as 
\begin{equation}\label{homezero}
  h_{\omega}(r_{*})=e^{B^-}
  \left(A_{1}e^{i\omega r_{*}}+A_{2}e^{-i\omega r_{*}}\right),  
\end{equation}
where we used the approximation:
\begin{equation}
    e^{B^-}\equiv e^{B(r_{*\, \rm min})}|_{k<0}
    \approx \frac{e^{\frac{1}{\sqrt{2}}} \sqrt{\left| k\right| }}{\sqrt[4]{2}}.
\end{equation}
The first term of the solution \eqref{homezero} is the ingoing mode which can be approximated for small frequencies as  
\begin{equation}\label{hin}
    h^{(in)}_{\omega}(r_{*})
    \approx A_{1}e^{B^-}\left(1+i\omega r_{*}\right).
\end{equation}

On the other hand the hydrodynamic expansion Eq.\eqref{HHydrodynamicExpansiont0} near the minimum of the potential is given by
\begin{align}
    h_{\omega}(r_{*})=&C_{1}+C_{0}\left(\frac{
    Z^{-}}{4 \left| k\right| ^{3/2}}+\frac{\sqrt{2} e^{\sqrt{2}} r_{*}}{\left| k\right| }\right)\nonumber\\
    &+\omega\left(C^{(1)}_{1}+C^{(1)}_{0}\left(\frac{Z^{-}}{4 \left| k\right| ^{3/2}}+\frac{\sqrt{2} e^{\sqrt{2}} r_{*}}{\left| k\right| }\right)\right)+\mathcal{O}(\omega^{2}),
\end{align}
where $Z^{-}=\left(2 \sqrt[4]{2}-4\ 2^{3/4}\right) e^{\sqrt{2}}-\sqrt{\pi } \text{erfi}\left(\sqrt[4]{2}\right)\approx -22.08$. 
Matching this equation with Eq. \eqref{hin} we obtain
\begin{align}
      C_{0}&=0,
      \qquad \qquad \qquad \qquad  C_{1}=A_{1}e^{B^-}, 
\\ 
      C^{(1)}_{0}&= A_1 \frac {i}{2}\,  k^2\,  e^{-B^-} ,
      \qquad \qquad C^{(1)}_{1}=-A_{1}\frac{iZ^{-}}{8}\sqrt{|k|}e^{-B^-}.
\end{align} 
Thus we can express the general solution for $h_{\omega}(r_{*})$, Eq. \eqref{eq:Bogoliubovk}, as 
\begin{align}
    h_{\omega}(r_{*})=&A_{1}\left[e^{B^-}
    -\frac{i\omega e^{-B^-}}{2}\left(\frac{Z^{-}\sqrt{|k|}}{4}-k^{2} \left(\frac{r_{*} e^{r_{*}^2 \left| k\right| }}{2 \left| k\right| }-\frac{\sqrt{\pi } \text{erfi}\left(r_{*} \sqrt{\left| k\right| }\right)}{4 \left| k\right| ^{3/2}}\right)\right)\right]+\mathcal{O}(\omega^{2}).
\end{align}
Considering the region near the boundary and changing the coordinate $r_*$ to  $r=\frac{1}{r_{*}}$, this solution can be rewritten as 
\begin{align}
    h_{\omega}(r)=&A_{1}\left[e^{B^-}
    -\frac{i\omega e^{-B^-}}{2}\left(\frac{Z^{-}\sqrt{|k|}}{4}-k^{2} \left(\frac{ \left| k\right| }{5r^{5}}+\frac{1}{3r^{3}}\right)\right)\right]+\mathcal{O}(\omega^{2}), 
\end{align}
and its derivative with respect to $r$ is 
\begin{align}
    h'_{\omega}(r)=
    -A_{1}\frac{i\omega e^{-B^-}}{2}k^{2} \left(\frac{ \left| k\right| }{r^{6}}+\frac{1}{r^{4}}\right)+\mathcal{O}(\omega^{2}).
\end{align}

Using the expression for the force given by \eqref{force} in the zero temperature case ($r_h =0$) one has 
\begin{equation}
\label{ForceT=0}
    F(t) = \frac{1}{2 \pi \alpha'} \left[X'(r_b, t)\; r_b^4 e^{\frac{ k}{r_b^2}}\right], 
\end{equation}
where $X(r,t)=h_{\omega}(r)e^{-i\omega t}$.

Therefore the admittance for $k<0$ is found to be 
\begin{align}
\label{AdmittanceT=0K<0}
    \chi(\omega)^{-}=\frac{2\pi\alpha'}{k^{2} \left(1+\frac{ \left| k\right| }{r_{b}^{2}}\right)e^{-\frac{|k|}{r^{2}_{b}}}}\left\{\frac{2ie^{2B^-}
    }{\omega }+\left[\frac{Z^{-}\sqrt{|k|}}{4}-k^{2} \left(\frac{ \left| k\right| }{5r_{b}^{5}}+\frac{1}{3r_{b}^{3}}\right)\right]\right\}. 
\end{align}
This means that the string has an effective tension $2\pi\alpha'/|k|$. We will comment more on this at the end of next subsection.  

%%%%%%%%%%%%%%%%%%%%%%
%%%%%%%%%%%%%%%%%%%%%%
%%%%%%%%%%%%%%%%%%%%%%

\subsection{The case $k>0$}

Here, we are going to solve Eqs. \eqref{h0} and \eqref{h1} in the case $k>0$. So the minimum of the potential \eqref{eq:potencialr*} 
is now 
\begin{align}
    V_{\rm min}=\left(2 \sqrt{2}-1\right) \left| k\right|. 
\end{align}

The hydrodynamic expansion analogous to Eq.\eqref{HHydrodynamicExpansiont0}  is
\begin{align}
\label{HHydrodynamicExpansionK>0}
    h_{\omega}(r_{*})=&C_{1}+C_{0}\left(\frac{\sqrt{\pi } \text{erf}\left(r_{*} \sqrt{\left| k\right| }\right)}{4 \left| k\right| ^{3/2}}-\frac{r_{*} e^{-r_{*}^2 \left| k\right| }}{2 \left| k\right| }\right)\nonumber\\
    &+\omega\left(C^{(1)}_{1}+C^{(1)}_{0}\left(\frac{\sqrt{\pi } \text{erf}\left(r_{*} \sqrt{\left| k\right| }\right)}{4 \left| k\right| ^{3/2}}-\frac{r_{*} e^{-r_{*}^2 \left| k\right| }}{2 \left| k\right| }\right)\right)+\mathcal{O}(\omega^{2}).
\end{align}
Close to the minimum of the potential this becomes
\begin{align}
\label{HHydrodynamicExpansionK>0Minimum}
    h_{\omega}(r_{*})=&C_{1}+C_{0}\left(\frac{Z'}{4 \left| k\right| ^{3/2}}+\frac{\sqrt{2} e^{-\sqrt{2}} r_{*}}{\left| k\right| }\right)\nonumber\\
    &+\omega\left(C^{(1)}_{1}+C^{(1)}_{0}\left(\frac{Z'}{4 \left| k\right| ^{3/2}}+\frac{\sqrt{2} e^{-\sqrt{2}} r_{*}}{\left| k\right| }\right)\right)+\mathcal{O}(\omega^{2}), 
\end{align}
where $Z^+=e^{-\sqrt{2}} \left(e^{\sqrt{2}} \sqrt{\pi } \text{erf}\left(\sqrt[4]{2}\right)-2 \left(\sqrt[4]{2}+2\ 2^{3/4}\right)\right)\approx -0.605$. 
In this region we can make the approximation 
\begin{equation}
e^{B^+} 
    \equiv e^{B(r_{*\, \rm min})}|_{k>0}
    =\frac{e^{-\frac{kr_{*}^{2}}{2}}}{r_{*}}\approx\frac{e^{-\frac{1}{\sqrt{2}}} \sqrt{\left| k\right| }}{\sqrt[4]{2}}.
\end{equation}

Following the discussion on the $k<0$ case of the previous subsection, the ingoing mode in the low frequency regime here can be written as 
\begin{equation}
    h^{(in)}_{\omega}(r_{*}) \approx A_1
    e^{B^+}
    \left(1+i\omega r_{*}\right). 
\end{equation}
Matching this expression with Eq.\eqref{HHydrodynamicExpansionK>0Minimum} we can write  
\begin{align}
      C_{0}&=0,
      \qquad \qquad \qquad \qquad C_{1}=A_{1}e^{B^+}, \\ 
      C^{(1)}_{0}&=A_{1}\frac{i  k^{2}}{2}e^{-B^{+}},
      \qquad \qquad C^{(1)}_{1}=-A_{1}\frac{1}{8} i \sqrt{\left| k\right| } Z^+e^{-B^{+}}.
\end{align}

Therefore, nearby the boundary with the $r={1}/{r_{*}}$ coordinate we have
\begin{align}
    h_{\omega}(r)=A_{1}\left[e^{B^{+}}
    -\frac{
    i\omega e^{-B^{+}}}{2}\left(\frac{1}{4}  \sqrt{\left| k\right| } Z^+- k^{2}\left(\frac{1}{3r^{3}}-\frac{ \left| k\right| }{5r^{5}}\right)\right)\right], 
\end{align}
and the derivative of this mode with respect to $r$ is
\begin{align}
    h'_{\omega}(r)=A_{1}\frac{
    i\omega e^{-B^{+}}k^{2}}{2}
    \left(\frac{\left| k\right| }{r^6}-\frac{1}{r^4}\right).
\end{align}

Following the steps of the case $k<0$ the admittance is given by
\begin{equation}
\label{AdmittanceT=0K>0}
     \chi(\omega)^{+}=\frac{2\pi\alpha'}{k^{2}\left(1-\frac{\left| k\right| }{r_{b}^2}\right)e^{\frac{k}{r_{b}^{2}}}}\left\{\frac{2ie^{2B^{+}}}{\omega }+
    \left[\frac{1}{4}  \sqrt{\left| k\right| } Z^+- k^{2}\left(\frac{1}{3r_{b}^{3}}-\frac{ \left| k\right| }{5r_{b}^{5}}\right)\right]
    \right\}. 
\end{equation}
This implies that the string has an effective tension coupled to the particle on the brane.  This result is analogous to the case $k<0$ obtained in the previous subsection. In the limit $|k|\ll r_b^2$ both results can be written as 
\begin{eqnarray}
\label{AdmittanceMP}
 \chi(\omega)^{\pm}&=&\frac{2\pi\alpha'}{k^{2}}\left\{\frac{2ie^{2B^{\pm}}}{\omega }+
    \left[\frac{1}{4}  \sqrt{\left| k\right| } Z^{\pm}- k^{2}\left(\frac{1}{3r_{b}^{3}}\mp \frac{ \left| k\right| }{5r_{b}^{5}}\right)\right]
    \right\}\nonumber\\ 
    &\approx& \frac{2\pi\alpha'}{|k|}\frac{2i}
    {\omega }. 
\end{eqnarray}
Note that these expressions for the admittance behave as a power-law of $|k|$ instead of an exponential law as in the finite temperature case, discussed in subsection \ref{amdtdif}. These expressions are singular in the limit $|k|\to 0$. So, this case will be considered separately in the next subsection. 

\subsection{The case $k=0$}

It is now interesting to analyse the limit $k\to 0$ to recover the pure AdS case. Since  the Eqs. \eqref{AdmittanceT=0K<0} and \eqref{AdmittanceT=0K>0} are singular in this limit, we should go back to Eq. \eqref{PoincareMotionEquation}, which for $k=0$  becomes
\begin{equation}
\label{MotionEquationRStarkigual0}
    \frac{d^{2}h_{\omega}(r_{*})}{dr_{*}^{2}}-\frac{2}{r_{*}}\frac{dh_{\omega}(r_{*})}{dr_{*}}+\omega^2  h_{\omega}(r_{*})=0.
\end{equation}

The general solution to this equation can be written as
\begin{equation}
    h_{\omega}(r_{*})=r^{\frac{3}{2}}_{*}\left(D_{1}H^{(1)}_{\frac{3}{2}}(\omega r_{*})+D_{2}H^{(2)}_{\frac{3}{2}}(\omega r_{*})\right), 
\end{equation}
where $D_1$ and $D_2$ are constants and $H^{(1)}_a$ and $H^{(2)}_a$ are the Hankel functions of first an second kind, respectively, of order $a$. Then, the admittance $\chi(\omega)=X(\omega)/F(\omega)$ can be calculated from the ingoing mode $H^{(1)}_{\frac{3}{2}}(\omega r_{*})$ at the IR ($r \to 0$) so that 
\begin{equation}
    \chi(\omega)=-\frac{4 \pi \alpha  H_{\frac{3}{2}}^{(1)}\left(\frac{\omega }{r_{b}}\right)}{r_{b}^2 \left(\omega  H_{\frac{1}{2}}^{(1)}\left(\frac{\omega }{r_{b}}\right)+3 r_{b} H_{\frac{3}{2}}^{(1)}\left(\frac{\omega }{r_{b}}\right)-\omega  H_{\frac{5}{2}}^{(1)}\left(\frac{\omega }{r_{b}}\right)\right)},
\end{equation}
which in the low frequency regime becomes 
\begin{equation}\label{limkigual0}
    \chi(\omega)=\frac{2\pi \alpha'}{r_{b}^2}\left(\frac{ i }{ \omega }-\frac{r_{b} }{ \omega ^2}\right).
\end{equation}
\noindent 
This expression agrees with \cite{Tong:2012nf, Edalati:2012tc, Giataganas:2018ekx} for the pure AdS space with $T=0$.

Comparing the imaginary parts of the admittances at zero temperature and $k \lessgtr 0$, Eq. \eqref{AdmittanceMP}, we see that the role played by $r_b^2$ in the pure AdS case, is played by the constant $|k|$  in our deformed metric setup.  Interestingly, $r_b$ is a UV  scale, while $k$ is an IR one.

\section{Conclusions} \label{sec:conclusions}

Here, in the Conclusions, we will summarize our achievements and results obtained within our deformed string/gauge model, by the introduction of an exponential factor $\exp{(k/r^2)}$ in the AdS$_5$ metric to study a holographic description of the Brownian motion. Our choice is based on the idea of  breaking the conformal invariance but keeping the Lorentz symmetry for the boundary theory instead of a Lifshitz scale or a hyperscaling violation  as was done, for instance, in Refs. \cite{Tong:2012nf, Edalati:2012tc, Giataganas:2018ekx}. Our geometric setup is interesting because it may help the  description of  random motion of a massive quark in the quark-gluon plasma \cite{Dudal:2018rki}.

Within our model we started studying the finite temperature scenario. In order to do this we have included a horizon function in the AdS$_5$ metric dealing with a deformed  AdS-Schwarzschild black hole which is dual to a boundary field theory at finite temperature. In this scenario we computed the string  energy for positive and negative $k$, as can be seen in Eqs. \eqref{eq:Estringk>0} and \eqref{eq:Estringk<0}, in agreement with Refs. \cite{deBoer:2008gu, Edalati:2012tc}, which also reproduce the pure AdS behavior (without deformation), as showed in Eq. \eqref{eq:EAdS}.  In section \ref{FDT} we have computed the admittance or linear response $\chi(\omega)$, Eq. \eqref{eq:chitemperature}, and soon after, computing the diffusion constant, presented in Eq. \eqref{diffusion}. Both results are compatible with the literature \cite{Giataganas:2018ekx, Dudal:2018rki}. It is worthy to mention that the sign of the constant $k$ seems to be irrelevant for the admittance behavior at high temperatures, as can be seen in Figure \ref{chi}.

In subsection \ref{displace} we have computed the mean square displacement $s^2_{{\rm reg}}(t)$, from which we have obtained the ballistic and diffusive regimes of Brownian motion. In the short time limit from our deformed string/gauge model we find  $s^2_{{\rm reg}}(t) \sim t^2$, Eq. \eqref{balfinal}, which is the ballistic regime, as expected. For the long time limit  we find $s^2_{\rm reg}(t) \sim D t$, Eq. \eqref{diffinal}, which is the diffusive regime \cite{Kubo1966}. Going further in the finite temperature scenario within our model,  in subsection \ref{teofd},  we have checked the fluctuation-dissipation theorem, as one can see in Eq. \eqref{fludis}. 

Our last discussion is related to the zero temperature scenario. In this study, the horizon function in Eq. \eqref{HF} is reduced to $f(r)=1$. Thus, the AdS deformed metric for $T=0$ can be written as in Eq. \eqref{adst0} and the equation of motion (EOM), given by Eq. \eqref{PoincareMotionEquation}, was solved in the hydrodynamic approximation. We obtained the solutions for $k \lessgtr 0$ and the corresponding admittances, Eqs. \eqref{AdmittanceT=0K<0} and \eqref{AdmittanceT=0K>0}. It is important to mention that the admittances for $T=0$ behave as a power-law of $|k|$ while for the finite temperature case it is an exponential law. 
It is also worthy to note that the admittances found here in the deformed AdS space are singular in the limit $|k|\to 0$ in opposition to the finite temperature case where this limit is smooth.

\begin{acknowledgments}
The authors would like to thank R\^{o}mulo Rougemont and for useful discussions. We also thank an anonymous referee for interesting suggestions to improve the text. N.G.C. is supported by  Conselho Nacional de Desenvolvimento Científico e Tecnológico (CNPq) and Coordenação de Aperfeiçoamento de Pessoal de Nível Superior (CAPES). H.B.-F. and C.A.D.Z. are partially supported by Conselho Nacional de Desenvolvimento Cient\'{\i}fico e Tecnol\'{o}gico (CNPq) under the grants Nos. 311079/2019-9 and  309982/2018-9, respectively.
\end{acknowledgments}

\bibliographystyle{apsrev4-1} 
  \bibliography{fluctuation}

\end{document}